\documentclass[showpacs,preprintnumbers,reprint,amsmath,amssymb,prl]{revtex4-1}

\usepackage{graphicx}
\usepackage{dcolumn}
\usepackage{bm}
\usepackage{multirow}

\graphicspath{{./Figures_reformat/}}

\begin{document}

\title{A Terradynamics of Legged Locomotion on Granular Media}
\author{Chen Li$^{1,2}$, Tingnan Zhang$^{1}$, and Daniel I. Goldman$^{1*}$}
\affiliation{$^1$School of Physics, Georgia Institute of Technology, Atlanta, Georgia 30332, USA\\
$^2$Department of Integrative Biology, University of California, Berkeley, California, 94720, USA\\
$^*$Corresponding author. E-mail: daniel.goldman@physics.gatech.edu}

\maketitle


\textbf{The theories of aero- and hydrodynamics predict animal movement and device design in air and water
through the computation of lift, drag, and thrust forces. Although models of terrestrial legged locomotion
have focused on interactions with solid ground, many animals move on substrates that flow in
response to intrusion. However, locomotor-ground interaction models on such flowable ground are
often unavailable. We developed a force model for arbitrarily-shaped legs and bodies moving
freely in granular media, and used this ``terradynamics'' to predict a small legged robot's locomotion
on granular media using various leg shapes and stride frequencies. Our study reveals a complex but
generic dependence of stresses in granular media on intruder depth, orientation, and movement
direction and gives insight into the effects of leg morphology and kinematics on movement.}

The locomotion of animals (1) and devices (2--4) emerges from the effective interaction of bodies and/or appendages with an environment. For flying in air and swimming in water, there is a history of theoretical predictive models (3--5) to describe the complex interactions between the locomotor and the surrounding fluids,
based on the fundamental equations for fluid flow, the Navier-Stokes equations. These models have not only allowed understanding of the movement of a variety of aerial and aquatic organisms (5) [such as bacteria and spermatazoa (6), insects (7), birds (8), and fish and whales (9)] and their functional morphology,
evolution, and ecology (9, 10), but also advanced the engineering design of aircraft (3), marine vehicles (4), and flying (11) and swimming (12) robots. For running and walking on ground, studies using solid ground such as
running tracks and treadmills have inspired general models (13, 14); building on these models, researchers
have begun to apply dynamical systems theory (15). In these studies, the leg-ground interaction was often approximated as a point contact on a rigid, flat, and nonslip ground (13--15).

\begin{figure}[b!]
\begin{centering}
\includegraphics[width=3.3in]{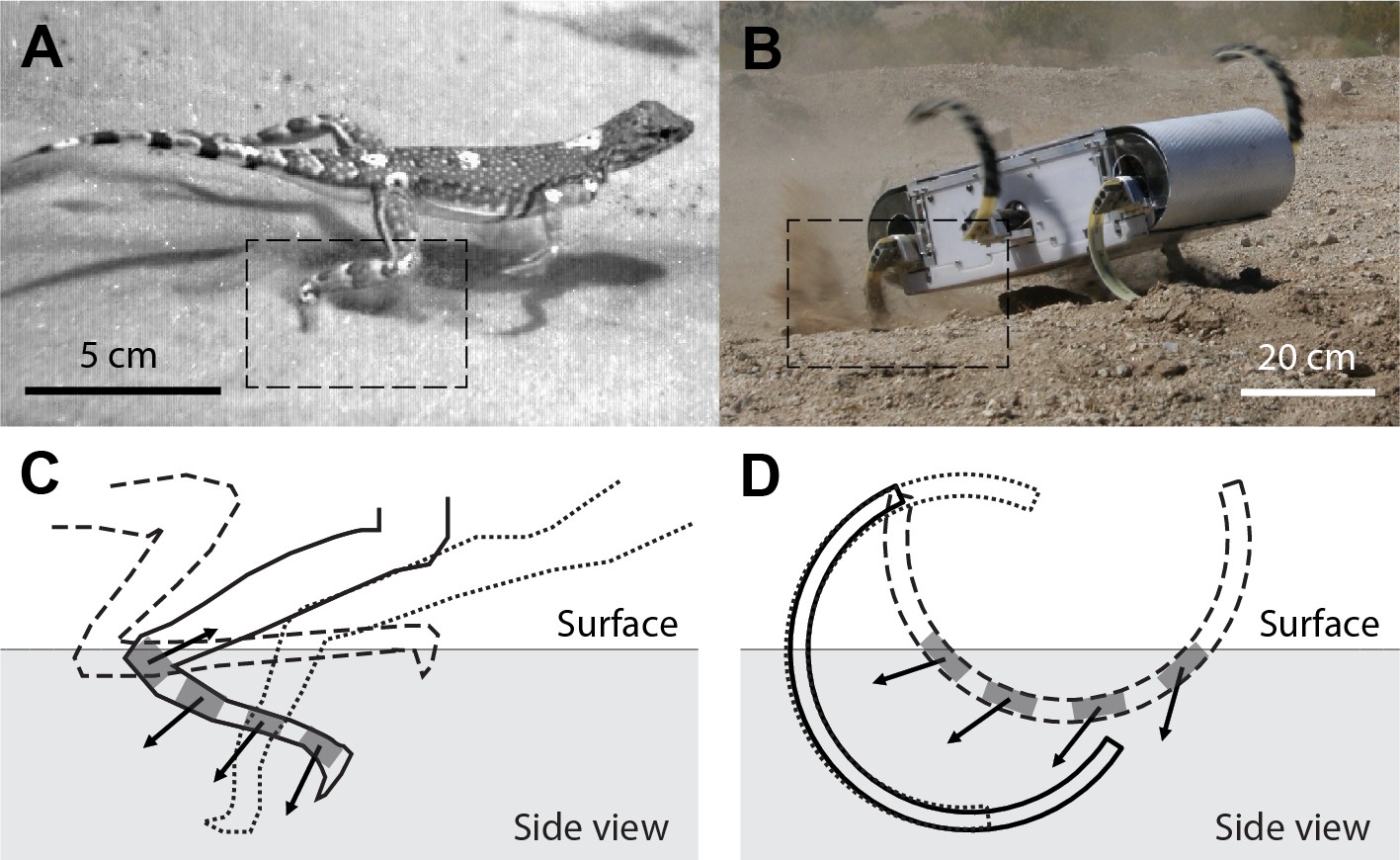}
\end{centering}
\caption
{
Examples of legged locomotion on flowable ground. \textbf{(A)} A zebra-tailed lizard running on sand (16). \textbf{(B)} A biologically inspired RHex robot (22) walking on dirt [Photo credit: Galen Clark Haynes, Aaron M. Johnson, and Daniel E. Koditschek, University of Pennsylvania]. Dashed boxes in (A) and (B) indicate the regions of leg-ground interaction shown in (C) and (D). Schematic of leg-ground interaction for \textbf{(C)} a hind foot of a zebra-tailed lizard (16) and \textbf{(D)} a c-leg of a RHex robot (21) during a step on granular media. Dashed, solid, and dotted tracings are leg positions at early, mid-, and late stance. Bars and arrows indicate local orientations and movement directions of leg elements. The gray area is the granular substrate.
}
\label{Motivation}
\end{figure}

Many small legged animals (16--19) [and increasingly robots (20--23)] face the challenges of moving on natural substrates such as sand (16, 17, 21), gravel (16, 20), rubble (20), soil (20, 22),mud (17, 20), snow (18, 20), grass (20, 22), and leaf litter (19, 20, 22), which, unlike solid ground, can flow during movement when a yield
stress is exceeded. The complexity of the interactions with such ¡°flowable ground¡± may rival or even exceed that during movement in fluids. For example, recent studies of legged animals (16) and robots (21) moving on granular media [collections of particles (24)] such as sand and gravel (Fig. 1, A and B) have demonstrated that at an instant of time during a step, each element of a leg moves through the substrate at a specific depth,
orientation, and movement direction, all of which can change over time (16, 21). Furthermore, the leg interacts with a material that can display both solid-like and fluid-like features (24) (Fig. 1, C and D). Compared to the theories of aero- and hydrodynamics, predictive models are less well developed for calculating forces and predicting legged locomotion on such flowable ground (16, 21). Research in the field of terramechanics (2)
has advanced the mobility of off-road vehicles on flowable ground such as sand and soil. These models were developed for large wheeled and tracked vehicles, which sink only slightly into the substrate (2, 25). Thus, in terramechanical models, interaction with the ground is approximated as the indentation of a horizontal, flat,
rectangular plate (2, 25). It was a breakdown of this flat-plate approximation, however, that led to overpredicted speeds for small vehicles such as the Mars rovers, whose small wheels have substantially curved ground contact interfaces (25). Because leg-ground interaction on flowable substrates is a more diverse, complex, and dynamic process (16, 21) than the flat-plate indentation, terramechanics is not expected to apply
to legged locomotors on flowable ground (2).

Granular materials such as sand and gravel (24), home to a variety of small desert animals (16, 17, 26), have proved to be a promising model substrate for studying legged locomotion on flowable ground (16, 21). Despite their diversity in particle size, shape, density, friction, polydispersity, and compaction (24), dry granular media are relatively simple as compared to media such as soil and mud, because granular particles interact purely through dissipative, repulsive contact forces and have no cohesion (24). In addition, the penetration
resistance of granular media can be repeatably controlled using laboratory apparatus such as an air-fluidized bed (16, 21, 26).

\begin{figure}[b!]
\begin{center}
	 \includegraphics[width=3.3in]{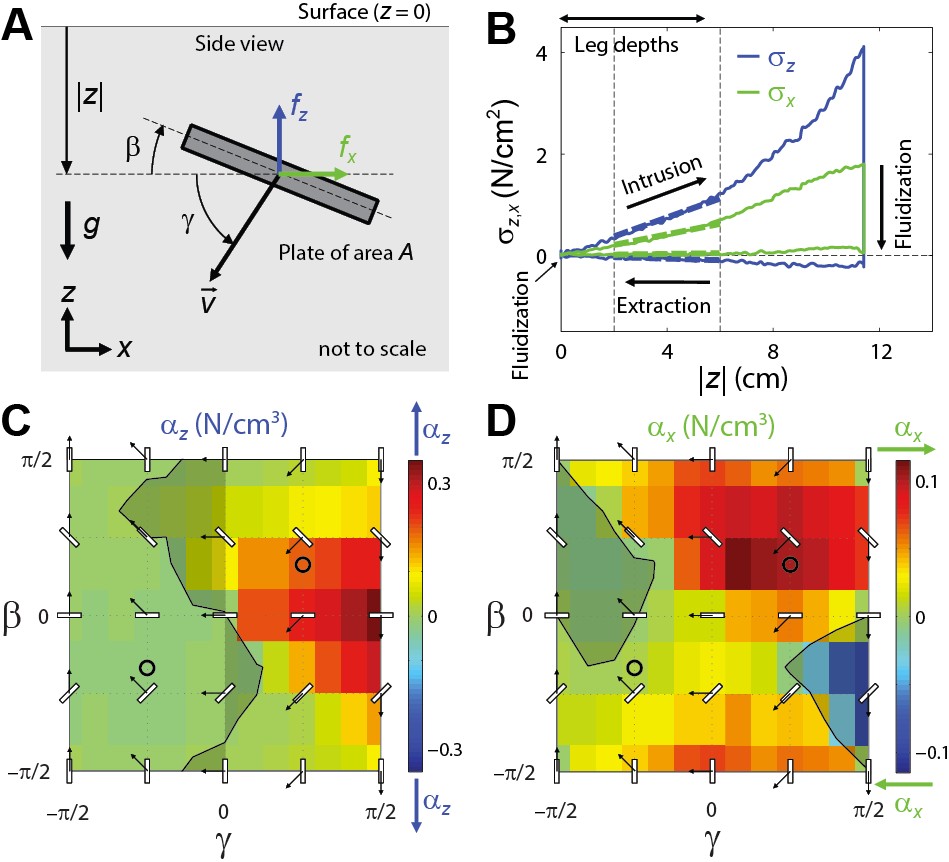}
\end{center}
\caption
{
Measurement of resistive forces in granular media in the vertical plane using a plate element (28). \textbf{(A)} Lift $f_z$ (blue arrow) and drag $f_x$ (green arrow) on a thin rigid plate of area A moving in granular media (gray area) in the vertical plane at speed $v = 1$ cm/s were measured as a function of the plate's depth $|z|$ below the surface, angle of attack $\beta$, and angle of intrusion $\gamma$. The granular media were fluidized (and then compacted when a closely packed state was prepared) using an air-fluidized bed (26) before each intrusion ($\gamma \geq 0$) and extraction ($\gamma \leq 0$). $g$ is gravitational acceleration. \textbf{(B)} Vertical (blue curve) and horizontal (green curve) stresses $\sigma_{z,x} = f_{z,x}/A$ versus $|z|$ for representative intrusion and extraction using $(\beta, \gamma) = \pm(\pi/6, \pi/4)$ (movie S1). Blue and green dashed lines are linear fits with zero intercept over intermediate depths at which the plate was fully submerged and far from the bottom of the container. Horizontal arrows on top indicate the range of leg depths in Figs. 3 and 4. \textbf{(C)} Vertical and \textbf{(D)} horizontal stresses per unit depth $\alpha_{z,x}$ [slopes of dashed fit lines in (B)] versus $\beta$ and $\gamma$. Plate schematics with arrows denote representative orientations and movement directions. Black curves indicate where $\alpha_{z,x} = 0$. The shaded areas indicate where $\alpha_z$ (or $\alpha_x$) is not opposing the plate's vertical (or horizontal) velocity. Circles indicate $\alpha_{z,x}$ from data shown in (B). Arrows above and below the color bars indicate directions of $\alpha_{z,x}$ for positive and negative values.
}
\label{RFTdevelopment}
\end{figure}

To begin to create a ``terradynamics'' that allows the prediction of legged locomotion on a flowable ground, we hypothesized that the net forces on a leg (or a body) moving in a granular medium in the vertical plane could be approximated by the linear superposition of resistive forces on infinitesimal leg (or body) elements.
Our hypothesis was inspired by our recent success in applying the methods of resistive force theory (6) to predict the forces and movement during the limbless locomotion of a lizard swimming in sand (26) and in describing the drag (26, 27) and lift (27) on simple objects moving in granular media at fixed depths. In these studies, the linear superposition was valid for intruders moving in granular media in the horizontal plane
at low enough speeds [for example, $\leq 0.5$~m/s for 0.3-mm glass particles (26)], where intrusion forces are dominated by particle friction (insensitive to speed) and non-inertial (26). However, it was unclear whether linear superposition could apply to legs (or bodies) of complex morphology and kinematics moving in the vertical plane.

To measure resistive forces for leg elements, we moved a thin rigid plate (of area A) in granular media in the vertical plane at 1 cm/s and measured lift $f_z$ and drag $f_x$ on the plate (in the continuously
yielding regime). We determined vertical and horizontal stresses $\sigma_{z,x} = f_{z,x}/A$ as a function of the plate's depth $|z|$ below the surface, angle of attack $\beta$, and angle of intrusion $\gamma$ (Fig. 2A and movie S1) (28). To test the generality of our resistive force model, we used three dry granular media of various particle size, shape, density, and friction, prepared into flat, naturally occurring, loosely and closely packed states (16, 21, 26) (supplementary text section 2, fig. S3, and table S1). Slightly polydispersed near-spherical glass particles 0.3 and 3 mm in diameter [covering the particle size range of natural dry sand ($\sim$0.1 to $\sim$1 mm) (29)] and rounded, slightly kidney-shaped poppy seeds (0.7 mm in diameter) allowed us to probe general principles for naturally occurring granular media of high sphericity and roundness [such as Ottawa sand (30)]. We discuss at the end of the paper possible effects of particle nonsphericity and angularity also found in many natural sands (30).

In all media tested, we observed that for all attack angles $\beta$ and intrusion angles $\gamma$, $\sigma_{z,x}$ were nearly proportional to depth $|z|$ when the plate was fully submerged and far from the bottom of the container (Fig. 2B and movie S1). This is because friction-dominated forces are proportional to the hydrostatic-like pressure in granular media. Therefore, we modeled the hydrostatic-like stresses as
\begin{equation}
\sigma_{z,x}(|z|, \beta, \gamma) =
\begin{cases}
\alpha_{z,x}(\beta, \gamma)|z| & \text{if $z < 0$}\\
0 & \text{if $z > 0$}
\end{cases}
\label{LinearFit}
\end{equation}
where $\alpha_{z,x}$ are vertical and horizontal stresses per unit depth (slopes of dashed fit lines in Fig. 2B). We found that in all media tested, $\alpha_{z,x}$ depended sensitively on both attack angle $\beta$ and intrusion angle $\gamma$ (Fig. 2, C and D, fig. S4, and additional data table S5). $\alpha_z$ (or $\alpha_x$) was opposing the plate's vertical (or horizontal) velocity for most but, counterintuitively, not all $\beta$ and $\gamma$ (exceptions are indicated by the shaded regions). For almost all attack angles $\beta$, $\alpha_{z,x}$ had larger magnitudes ($|\alpha_{z,x}|$) for intrusion angle $\gamma \geq 0$ than for $\gamma \leq 0$; i.e., it was harder to push the plate into granular media than to extract it. For all intrusion angles $\gamma$ except $\gamma = \pm \pi/2$, $|\alpha_{z,x}|$ were asymmetric to attack angles $\beta = 0$ and $\beta = \pm \pi/2$; i.e., only when the plate moved vertically were stress magnitudes the same for vertically or horizontally mirrored orientations (e.g., $\beta = \pm \pi/6$). These asymmetries are a result of gravity breaking symmetry in the vertical plane and differ from the case in the horizontal plane (26). Our resistive force measurements are an advance from previous force models based on the flat-plate approximation used in many terramechanical models (2, 25), which capture only the dependence of stresses on intruder depth, but not on its orientation or movement direction (supplementary text section 1 and fig. S1). Despite their different magnitudes and subtle differences in shape, the overall profiles of stresses (per unit depth) $\alpha_{z,x}(\beta, \gamma)$ were similar for all media tested. Furthermore, these stress profiles could be approximated (to the first order) by a simple scaling of generic stress profiles (supplementary text section 3, figs. S6 and S7, and tables S2 and S3).

\begin{figure}[b!]
\begin{center}
	 \includegraphics[width=3.3in]{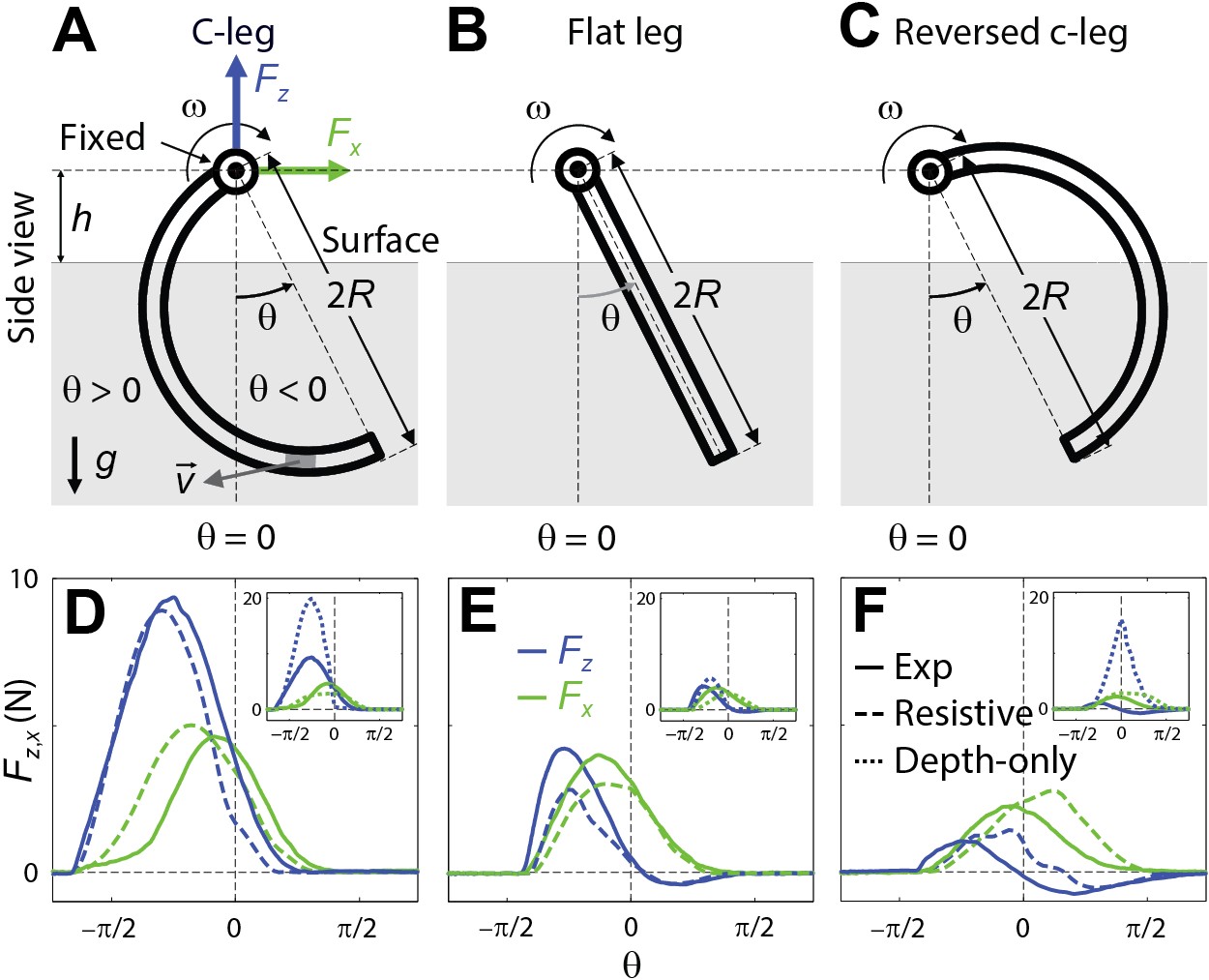}
\end{center}
\caption
{
The resistive force model predicts forces on intruders of complex morphology and kinematics moving in granular media (28). Three thin rigid model legs of different geometries, \textbf{(A)} a c-leg, \textbf{(B)} a flat leg, and \textbf{(C)} a reversed c-leg, were rotated about a fixed axle at a hip height h through granular media (gray area) in the vertical plane at an angular velocity $\omega$, generating leg speeds of $v \sim 1$ cm/s, and net lift $F_z$ (blue) and thrust $F_x$ (green) were measured as a function of leg angle $\theta$ (movie S2). All three legs had identical maximal length $2R$ from the axle. $g$ is gravitational acceleration. (\textbf{D} to \textbf{F}) $F_{z,x}$ versus $\theta$ on the three legs measured in the experiment (solid curves) and predicted by the resistive force model (dashed curves) using Eq. 2 (movie S3). Insets: $F_{z,x}$ versus $\theta$ from experiment (solid curves) versus predicted (dotted curves) using Eq. 2 and previous force models in which stresses depended only on depth (supplementary text section 1 and fig. S2).
}
\label{RFTvalidation}
\end{figure}

We next tested our hypothesis that forces on a complex intruder moving in granular media in the vertical plane could be approximated by the linear superposition of forces on all intruder elements. We measured the net lift $F_z$ and thrust $F_x$ on thin rigid model legs rotating about a fixed axle [simulating a tethered body (7, 11)] through granular media in the vertical plane at $\sim$1 cm/s (Fig. 3 and movie S2) (28). We then compared
them to predictions from the resistive force model by the integration of stresses over the legs (movie S3)
\begin{equation}
\begin{aligned}
F_{z,x} = \int\limits_S \! \sigma_{z,x}(|z|_s, \beta_s, \gamma_s) \, dA_s = \int\limits_S \! \alpha_{z,x}(\beta_s, \gamma_s) |z|_s \, dA_s
\end{aligned}
\label{Integration}
\end{equation}
where $S$ is the leading surface of the leg; $dA_s$, $|z|_s$, $\beta_s$, and $\gamma_s$ are the area, depth below the surface, angle of attack, and angle of intrusion of infinitesimal leg elements; and $\alpha_{z,x}(\beta_s, \gamma_s)$ are element stresses per unit depth (interpolated from data in Fig. 2, C and D). To test the robustness of our force model, we used three model legs of different geometries [with the same maximal leg length $2R$ (28)]: a RHex robot's c-leg (21, 22), a flat leg, and a reversed c-leg (Fig. 3, A to C). In model calculations, each leg was divided into 30 elements.

In all media tested (Fig. 3,D to F, and fig. S5), we observed that for all three legs, the measured net lift and thrust $F_{z,x}$ as a function of leg angle $\theta$ (solid curves) were asymmetric to the vertical downward direction ($\theta = 0$), and were larger during intrusion ($\theta \leq 0$) than during extraction
($\theta \geq 0$). Peak $F_{z,x}$ were largest on the c-leg and smallest on the reversed c-leg. The reversed c-leg experienced significant negative lift (suction force, $F_z < 0$) during extraction. For all media tested, our resistive force model predicted $F_{z,x}$ versus $\theta$ for all three legs (dashed curves), capturing both the magnitudes and asymmetric profiles. The relative errors of peak forces between data and model predictions were within 10\% for the c-leg in four media tested, and within 33\% for all three legs in all media tested. The accuracy of our resistive force model was significantly better than that of previous force models in which stresses depended only on depth (Fig. 3, D to F, insets; supplementary text section 1 and fig. S2). Furthermore, our resistive force model revealed that the c-leg generated the largest forces, because its morphology allowed leg elements to not only reach deeper depths but also access larger stress regions in Fig. 2, C and D (particularly for elements at large depths).

Our discovery of the insensitivity of the stress profiles to particle properties (fig. S4) has practical benefits: For granular media of near-monodispersed, near-spherical, rounded particles, as an alternative to measuring $\alpha_{z,x}$ for all attack angles $\beta$ and intrusion angles $\gamma$ in the laboratory, one can simply perform a single measurement [of $\alpha_z(0, \pi/2)$, using a horizontal plate penetrating vertically downward] to infer all $\alpha_{z,x}(\beta, \gamma)$ by a scaling routine (supplementary text section 3 and fig. S9) and predict forces (with a small loss in accuracy for the c-leg and the flat leg, but a larger loss in accuracy for the reversed c-leg, fig. S8).

We tested the ability of our resistive force model to predict legged locomotion. We chose to study the locomotor performance (speed) of a small RHex-like robot (22) (Fig. 4A, top, and movie S4) moving on granular media (28). The robot's six legs rotated nearly entirely in the vertical plane during locomotion, and its small
size ensured that leg intrusion speeds were low enough for particle inertia to be negligible. We chose poppy seeds as the test granular medium, because the grains were both small enough be prepared in our fluidized bed track (21) and large enough to not jam the robot's motor and gear trains. The robot's legs had a similar friction coefficient with poppy seeds to that of the model legs and were sufficiently rigid so that they experienced negligible bending during movement (28).

Unlike the sand-swimming lizard, which moves within granular media quasistatically (thrust and drag are always roughly balanced) (26), legged locomotion on the surface of granular media is dynamic (forces are not always instantaneously balanced). As a result, the resistive force theory (which solves for speed by balancing forces)
(6, 26) cannot be directly applied. Thus, to use our resistive force model to calculate robot speed, we developed a three-dimensional multibody dynamic simulation of the robot (Fig. 4A, bottom) (28). The simulated robot had the same body and leg morphology and used the same alternating tripod gait as the actual robot and had its motion constrained in the vertical plane. We divided each body plate and leg into 30 elements. The velocity $\vec{v}$ and angular velocity $\vec{\omega}$ of the simulated robot's body were calculated by

\begin{equation}
\begin{cases}
\vec{v}(t + dt) = \vec{v}(t) + \frac{\vec{F}}{m}dt\\
\vec{\omega}(t + dt) = \vec{\omega}(t) + \frac{\vec{N}}{I}dt\\
\end{cases}
\label{Dynammics}
\end{equation}
where $\vec{F}$ and $\vec{N}$ are the sum of net forces and torques on all the six legs and the body exerted by the granular medium, calculated from our resistive force model by the integration of stresses over each leg and the body using Eq. 2; $m$ and $I$ are the robot's mass and moment of inertia; and $t$ and $dt$ are time and time step. To test the robustness of our resistive force model and simulation, we used legs of seven geometries with different curvatures $1/r$ (given maximal leg length $2R'$) (Fig. 4C, inset) and varied stride frequency $f$ to up to 5 Hz (28).

\begin{figure}[b!]
\begin{center}
	 \includegraphics[width=3.3in]{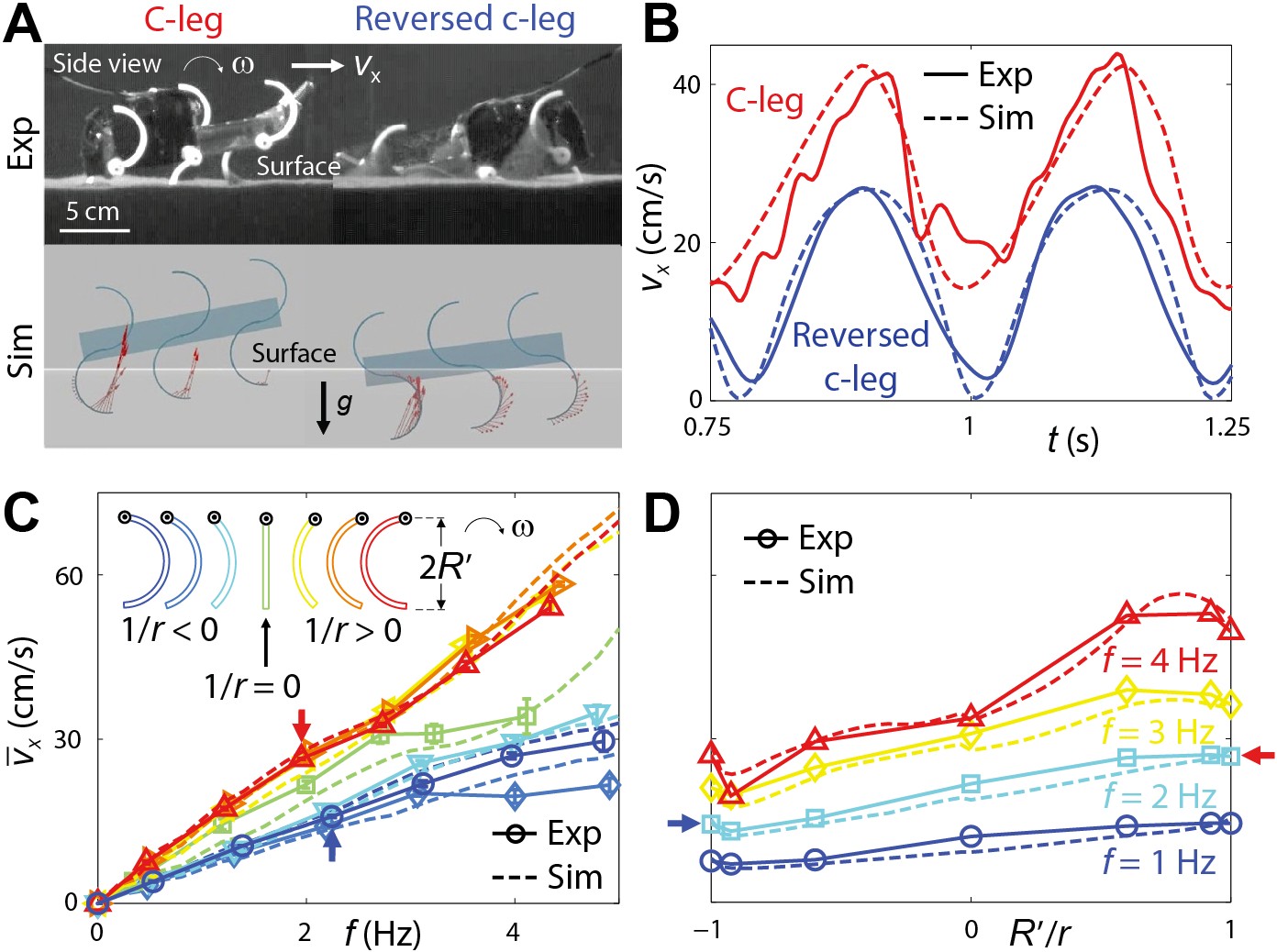}
\end{center}
\caption
{
A multibody dynamic simulation using the resistive force model predicts legged locomotion on granular media (28). \textbf{(A)} Side views of a small RHex-like robot (movie S4) at mid-stance during locomotion on granular media, using c-legs (left) and reversed c-legs (right) in the experiment (top, movie S5) and simulation (bottom,movie S6). Arrows in the simulation indicate element forces on one tripod of legs. $g$ is gravitational acceleration. \textbf{(B)} Forward speed $v_x$ versus time $t$ from two representative runs using c-legs (red, stride frequency $f = 2.0$ Hz, curvature $1/r = 1/R'$) and reversed c-legs (blue, $f = 2.2$ Hz, $1/r = -1/R'$). \textbf{(C)} Average forward speed $\overline{v}_x$ versus $f$ using legs of seven curvatures $1/r$ transitioning from reversed c-legs to c-legs (inset), where $r$ is the radius of curvature, $2R'$ is the maximal length of the robot legs, and the minus sign denotes reversed legs. \textbf{(D)} $\overline{v}_x$ versus $1/r$ at $f =$ 1, 2, 3, and 4 Hz. In (B) to (D), solid and dashed curves indicate the experiment and simulation, respectively. Error bars in (C) denote $\pm 1$ SD ($\ll 1$ cm/s in the simulation). Experimental data in (D) are interpolated from those in (C) (hence no error bars). Red and blue arrows in (C) and (D) indicate averages from data shown in (B).
}
\label{LocomotorPeformance}
\end{figure}

We observed similar robot kinematics (Fig. 4A) and forward speed $v_x$ versus time $t$ (Fig. 4B) in both the experiment (movie S5) and simulation (movie S6). The robot moved faster and penetrated its legs less deeply during stance using c-legs (Fig. 4A, left; Fig. 4B, red) than using reversed c-legs (Fig. 4A, right; Fig. 4B, blue). Average forward speed $\overline{v}_x$ increased with stride frequency $f$ for legs of all curvatures $1/r$ and was lower at any $f$ using legs of negative curvatures than using legs of positive curvatures (Fig. 4, C and D). The agreement between experiment and simulation in $\overline{v}_x(f, 1/r)$ was remarkable: Errors were within 20\% for 90\% of the $f$ and $1/r$ tested and within 35\% for all the $f$ and $1/r$ tested. Simulation using our scaling routine also achieved reasonable accuracy (supplementary text section 5 and fig. S13). This was an improvement over simulation using previous force models in which stresses depended only on depth (fig. S13). Our resistive force model and simulation revealed that the robot moved faster using c-legs than using reversed c-legs, because whereas the c-legs penetrated less deeply, their elements accessed larger stress regions in Fig. 2, C and D, resulting in larger leg lift (fig. S14) and smaller body drag. Our model and simulation also allowed the prediction of ground reaction forces on granular media(Fig. 4A, red arrows, and fig. S13), which would be difficult to measure otherwise. Furthermore, our model and simulation predicted that using arc-like legs (given maximal leg length $2R'$ ) of an optimal curvature of $1/r = 0.86/R'$, the robot would achieve maximal speed of $\overline{v}_x = 72$ cm/s ($\approx 5$~body length/s) at $5$ Hz. Our approach affords significant reduction in the computational time needed to model movement on granular media. For example, relative to our multiparticle discrete element method (DEM) simulation of movement on granular media (23, 27), our simulation using the resistive force model can achieve a factor of $10^6$ in speed-up (e.g., 10 s versus 30 days using DEM to simulate 1 s of locomotion on a granular bed of $5 \times 10^6$ poppy seeds).

We close with a brief discussion of the limitations of our model.We tested the predictive power of our scaling routine (supplementary text section 3 and fig. S9) for two highly polydispersed, nonspherical, highly angular natural sands (supplementary text section 4, figs. S10 to S12, and table S4). We found that the model accuracy
for natural sands was slightly worse than found in the glass spheres and poppy seeds (for example, 35\% versus 20\% error in peak $F_z$ for a rotating c-leg). As was the case for the near-spherical granular media tested, the functional forms of forces on the c-leg and the flat leg were still well captured by our scaling routine. Furthermore, the overestimation was not affected by reducing the polydispersity of the natural sand. This suggests that the nonsphericity and angularity of natural sand particles (30) may be the cause of this overestimation, which may require additional model fitting parameters and scaling factors. Our model
is intended for dry sand [$\sim$0.1 to $\sim$1 mm in particle diameter (29)] and may not work for dry cohesive powder ($\leq \sim$0.01 mm) (31). We do not expect our model to work if the particle size approaches a characteristic length of the locomotor (for example, $\sim$1-cm particles for our robot of $\sim$1-cm foot size), so that the continuum assumption breaks down and particles effectively become ¡°boulders.¡± We also do not expect our model to capture wet,
cohesive flowable media such as soil and mud.

We have developed a new approach to predicting legged locomotion on granular media. This terradynamics relies on new resistive force measurements and linear superposition (6, 26, 27). The general profiles of these resistive force measurements are insensitive (other than magnitudes) to a variety of granular media composed of slightly polydispersed, approximately spherical, rounded particles. Our terradynamics may not be limited to legged locomotion, because the integration of stresses should in principle work for devices of other morphology and kinematics, such as wheels, tracks, and earth movers moving on granular media (2, 25). For the particle types tested here, an important addition to our model would be to capture three-dimensional effects (16) and
spatial and temporal variation in compaction (21) and slopes (17), and test its validity in the high speed
``inertial fluid'' regime (when leg intrusion speed is $\geq 1$ m/s, at which particle inertia dominates forces) (23). Our resistive force model also provides opportunities to test and develop new physics theories of dense granular flow (32). Finally, we envision that, in concert with aero- and hydrodynamics (3--12), a general terradynamics
of complex ground will not only advance understanding of how animals move (1) at present (5--10, 13--19, 26) and in the past (17, 33), but also facilitate the development of robots with locomotor capabilities approaching those of organisms (11, 12, 20--23).





\begin{flushleft}
\textbf{Acknowledgements}
\end{flushleft}

We thank Y. Ding, P. Umbanhowar, N. Gravish, G. Meirion-Griffith, S. Sharpe, H. Komsuoglu, D. Koditschek, and R. Full for discussions; J. Shen for assistance with robot modification; P. Masarati for multibody dynamic simulator support; S. Sharpe for measuring the angle of repose of 3-mm glass spheres and assistance with
photography; P. Umbanhowar and H. Marvi for natural sand collection; and all the members of the Complex
Rheology And Biomechanics Lab at Georgia Tech for general assistance. This work was supported by the Burroughs
Wellcome Fund, the Army Research Laboratory Micro Autonomous Systems and Technology Collaborative Technology
Alliance, the Army Research Office, and the NSF Physics of Living Systems program. C.L. was partially supported by a Miller Research Fellowship from the Miller Institute for Basic Research in Science of the University of California, Berkeley. The authors declare that they have no competing interests. C.L. designed the study, performed resistive force measurements, and performed robot experiments; C.L. and T.Z. performed model calculations; T.Z. performed robot simulation; D.I.G. oversaw the study; and C.L. and D.I.G. wrote the paper.

%
%
%
%
%
%
%
%

\begin{flushleft}
\textbf{Citation Information}
\end{flushleft}

Chen Li, Tingnan Zhang, Daniel I. Goldman, A Terradynamics of Legged Locomotion on Granular Media, \emph{Science} \textbf{339}, 1408--1411 (2013), DOI: 10.1126/science.1229163

\renewcommand{\thefigure}{S\arabic{figure}}
\renewcommand{\thetable}{S\arabic{table}}
\renewcommand{\theequation}{S\arabic{equation}}

%
%
%
%
%
%
%
%
%
%
%
%
%
%
%

\setcounter{figure}{0}
\setcounter{table}{0}
\setcounter{equation}{0}

\clearpage

\begin{flushleft}
\textbf{Materials and Methods}
\end{flushleft}

\begin{flushleft}
\underline{Force measurements}
\end{flushleft}

We used aluminum to construct the plate element (area $A = 3.81 \times 2.54$~cm$^2$, thickness $= 0.64$~cm) and model legs (maximal length $2R = 7.62$~cm, width $= 2.54$~cm, thickness $= 0.64$~cm). We measured the friction coefficient $\mu$ between aluminum and poppy seeds to be $0.40$ (later we constructed robot legs using plastic of a similar friction coefficient with poppy seeds, $0.36$), by placing an aluminum plate on a wooden plate bonded with a single layer of poppy seeds, increasing the slope of the wooden plate from zero, and examining the angle $\xi$ at which the aluminum plate began to slide. Thus $\mu =$ tan$\xi$. The length and width of both the plate element and model legs were $\sim 10$ times the particle diameter, ensuring that the granular media could be approximated as a continuum.

Before each force measurement, we used an air fluidized bed ($24 \times 22$~cm$^2$ surface area) (26) to prepare the granular media ($15$~cm deep) to a well-defined compaction (see table~\ref{GranularMediaUsed} for the volume fractions of closely and loosely packed states of the granular media tested). Air flow was turned off during force measurements. We used a $6$ degree-of-freedom robotic arm (CRS Robotics) to move the plate element and rotate the model legs. We used a $6$-axis force and torque transducer (ATI Industrial Automation) mounted between the intruder and the robotic arm to measure forces to a precision of $0.05$~N at a sampling frequency of $100$~Hz. We performed all the force measurements at low speeds ($\sim 1$~cm/s) to ensure that particle inertia was negligible, and in a vertical plane at the middle of the air fluidized bed and far from the sidewalls (distance $> 3$~cm) to minimize boundary effects.

In the plate element intrusion experiment, we attached the plate to the force and torque transducer via a supporting rod and an adjustable mount with which attack angle $\beta$ could be varied. We varied intrusion angle $\gamma$ by adjusting the trajectory of the robotic arm. For each combination of $\beta$ and $\gamma$, we separately measured the forces on the supporting rod and mount moving in the granular media without the plate, and subtracted them to obtain forces exerted by the granular media on the plate alone.

During each test session, we first prepared the granular media while the plate was above the surface. We then moved the plate (oriented at attack angle $\beta$) downward to the surface (depth $|z| = 0$), paused it for 2 seconds, and then intruded it into the granular media along intrusion angle $\gamma$. After intrusion was finished, we prepared the granular media again, and extracted the plate along the same path. This gave us measurements of stresses $\sigma_{z,x}$ for both $\pm(\beta, \gamma)$. For horizontal movements ($\gamma = 0$), $\sigma_{z,x}$ were nearly constant when the plate was far from the container sidewalls, and we obtained $\alpha_{z,x}$ by fitting Eq.~1 to averages of $\sigma_{z,x}$ in the steady state regions at three depths ($|z| = 2.54$~cm, $5.08$~cm, and $7.62$~cm). We measured $\alpha_{z,x}$ for $\gamma$ within $[-\pi/2, \pi/2]$, and determined $\alpha_{z,x}$ for $\gamma$ within $[-\pi, -\pi/2]$ and $[\pi/2, \pi]$ by symmetry:
\begin{equation}
\begin{cases}
\alpha_z(\beta, \gamma) = \alpha_z(-\beta, -\pi-\gamma) & \text{if $-\pi \leq \gamma \leq -\pi/2$}\\
\alpha_z(\beta, \gamma) = \alpha_z(-\beta, \pi-\gamma) & \text{if $\pi/2 \leq \gamma \leq \pi$}
\end{cases}
\end{equation}
\begin{equation}
\begin{cases}
\alpha_x(\beta, \gamma) = - \alpha_x(-\beta, -\pi-\gamma) & \text{if $-\pi \leq \gamma \leq -\pi/2$}\\
\alpha_x(\beta, \gamma) = - \alpha_x(-\beta, \pi-\gamma) & \text{if $\pi/2 \leq \gamma \leq \pi$}
\end{cases}
\end{equation}

In the model leg rotation experiment, we rotated the model legs at an angular velocity $\omega = 0.2$~rad/s at a hip height $h = 2$~cm within leg angle $-3\pi/4 \leq \theta \leq 3\pi/4$ (at the beginning and end of which all three legs tested were fully above the granular surface), where leg angle $\theta$ was defined as the angle sweeping from the vertical downward direction to the direction along which leg length was maximal. For each model leg, we separately measured the forces due to the weight of the model legs rotating in the air, and subtracted them to obtain the forces on the model legs exerted by the granular media during rotation.

Due to the high repeatability of our fluidized bed and robotic arm, we found that for all media tested, run-to-run variation in $\alpha_{z,x}$ for fixed $\beta$ and $\gamma$ was always within $0.005$~N/cm$^3$ at any given depth; thus we only performed one trial for each combination of $\beta$ and $\gamma$. All the stresses were calculated in the regions where the plate was far from the container boundaries (distance $> 6$~cm). We also confirmed that for low enough speeds, intrusion forces in granular media were insensitive to speed (for example, in loosely packed poppy seeds, at $v = 1$~m/s, force only increased by less than $20\%$ from that at $v = 1$~cm/s); thus, particle inertia was negligible.\\

\begin{flushleft}
\underline{Robot experiments}
\end{flushleft}

We built our robot (body length $= 13$~cm, body mass $= 150$~g) by modifying a small commercially available robot (RoboXplorer, Smart Lab). The robot had similar morphology and kinematics as a RHex robot, with a rigid body and six legs performing 1 degree-of-freedom rotation in an alternating tripod gait. We substituted the stock motor with a stronger one (RadioShack Super Speed $9$--$18$~VDC Hobby Motor, Model \# 273-256) and modified the gear trains (gear ratio: 47 revolutions in the motor transmits into 1 rotation of the legs). These changes increased maximal stride frequency $f$ to $5$~Hz. We removed the external body shell to reduce weight and the belly area (to $13 \times 2$~cm$^2$). This reduced drag on the belly during locomotion on granular media.

We used $3$-D printing to make custom robot legs. All the legs had the same maximal length $2R' = 4.1$~cm, width $= 1.0$~cm, and thickness $= 0.3$~cm, but different curvatures $1/r = [-1, -0.92, -0.60, 0, 0.60, 0.92, 1]/R'$. The ABS plastic used to fabricate the legs had a similar friction coefficient with poppy seeds ($0.36$) to that of aluminum with poppy seeds ($0.40$). The leg width and length were $\sim 10$ times the particle diameter, allowing the granular media to be approximated as a continuum. We ensured that the legs had large enough stiffness and moved like rigid bodies ($< 5\%$ deformation) during locomotion.

We tuned the center of mass of the robot to overlap with the geometric center of the body by adding mass to the lighter end of the robot. We measured the masses, dimensions, and relative positions of all robot body and leg parts, and calculated the moment of inertia of the robot to be $I = 2.08 \times 10^3$~g~cm$^2$ about the pitch axis through the center of mass. We powered the robot by an external power supply (Power Ten Inc.) to ensure constant voltage during trials, and adjusted voltage to vary stride frequency $f$ between trials.

Before each trial, we used an air fluidized bed track ($200 \times 50$~cm$^2$ surface area) to prepare the granular media ($12$~cm deep) to a well-defined compaction, using methods similar to those in (21). We used two synchronized high-speed cameras (X-PRI, AOS Technologies) to capture top and side views of the robot's locomotion at $500$~frame/s. We measured $f$ from the side view, and measured forward speed $v_x$ from the top view by digitizing a high contrast marker placed near the center of mass. For each $f$ and $1/r$, we performed three trials and reported mean $\pm$~s.d. for average forward speed $\overline{v}_x$ in experiment.\\

\begin{flushleft}
\underline{Robot simulation}
\end{flushleft}

We used a multibody dynamic simulator, MBDyn (34), to create a simulation of the robot locomotion on granular media. MBDyn features a full three-dimensional simulation with $6$ degrees of freedom ($3$ translations and $3$ rotations). We constructed the robot body using $3$ rigid plates (the front, rear, and belly surfaces) and $6$ rigid legs. We constrained the robot body movement within the vertical plane, with fore-aft and dorso-ventral translations and pitch, and allowed the legs to only rotate about their axles perpendicular to the vertical plane. For each time step, we calculated the depth $|z|_s$, attack angle $\beta_s$, and intrusion angle $\gamma_s$ for each element to determine element stresses $\sigma_{z,x}$ (using Eq.~1 and $\alpha_{z,x}$ interpolated from the data in Fig.~2, C and D). We then summed forces on all elements to obtain net forces $F_{z,x}$ using our resistive force model by Eq.~2, and calculated the body dynamics by Eq.~3. We found that dividing each body plate and leg into fewer elements could further increase simulation speed at the cost of model accuracy.

In simulation tests, we varied curvature $1/r$ between $-1/R' \leq 1/r \leq 1/R'$ in increments of $0.04/R'$, and varied stride frequency $f$ between $0 \leq f \leq 5$~Hz in increments of $0.2$~Hz. For each $f$ and $1/r$, we performed three trials using different initial conditions with a phase difference of $2\pi/3$ for each tripod. We found that this resulted in variation in average forward speed $\overline{v}_x$ of $\ll 1$~cm/s; thus, we reported only the means of $\overline{v}_x$ in simulation. We confirmed that at the maximal stride frequency tested ($5$~Hz) the leg speeds were $< 1$~m/s averaged over a stance, allowing particle inertia to be negligible.

\clearpage

\begin{flushleft}
\textbf{Supplementary Text}
\end{flushleft}

This supplementary text contains 5 sections:

In section 1, we review the depth-only vertical penetration and horizontal drag force models. Based on the flat-plate approximation, we then calculate stresses per unit depth and net forces on rotating legs using these depth-only force models, and compare with our resistive force measurements and model predictions.

In section 2, we present our resistive force measurements and model calculations for a variety of granular media tested, which have different particle size, density, friction, and compaction; from these results we discover that the stress profiles are generic.

In section 3, we develop a scaling routine to capture the similar stress profiles observed for the variety of granular media, by scaling generic stress profiles determined from averages of fits to the measured stress profiles. This provides a practical means to easily apply our resistive force model.

In section 4, we test the ability of our scaling routine to predict forces on intruders moving in natural sands, and discuss limitations of our model for natural sands.

In section 5, we compare the predictive accuracy for locomotor performance (speed) of the robot using our resistive force model to that using our scaling routine and that using the depth-only force models.


\begin{flushleft}
1. Depth-only force models and their limitations
\end{flushleft}

Previous studies of intrusion forces in granular media focused on simple intruders (e.g. plates, rods, and spheres) moving with simple kinematics (e.g. vertical penetration and horizontal drag). For example, the vertical force $f_z$ on a horizontal plate element ($\beta = 0$) moving vertically ($\gamma = \pm \pi/2$) in granular media was observed to be proportional to the plate's depth $|z|$ and area $A$ (21, 35):
\begin{equation}
f_z = \alpha_z(0,\textrm{sgn}(\dot{z})\pi/2)|z|A\\
\label{VerticalPenetrationForceModel}
\end{equation}
Similarly, the horizontal force $f_x$ on a vertical plate element ($\beta = \pm \pi/2$) moving horizontally ($\gamma = 0$) in granular media was proportional to plate depth $|z|$ and plate area $A$ (36--38):
\begin{equation}
f_x = \alpha_x(\pi/2,0)|z|A\\
\label{HorizontalDragForceModel}
\end{equation}
where $\dot{z}$ is the velocity of the plate in the vertical direction, and $\alpha_z(0,\textrm{sgn}(\dot{z})\pi/2)$ and $\alpha_x(\pi/2,0)$ are vertical and horizontal stresses per unit depth for $(\beta, \gamma) = (0, \pm \pi/2)$ and $(\pi/2, 0)$ (determined from measurements in Fig.~2, C and D).

Both the vertical penetration and horizontal drag force models only account for the dependence of stresses on the intruder's depth ($|z|$), but not the dependence on its orientation (attack angle $\beta$) or movement direction (intrusion angle $\gamma$). Hereafter we refer to these force models as the ``depth-only force models''.



\begin{figure}[t!]
\begin{center}
	 \includegraphics[width=3.3in]{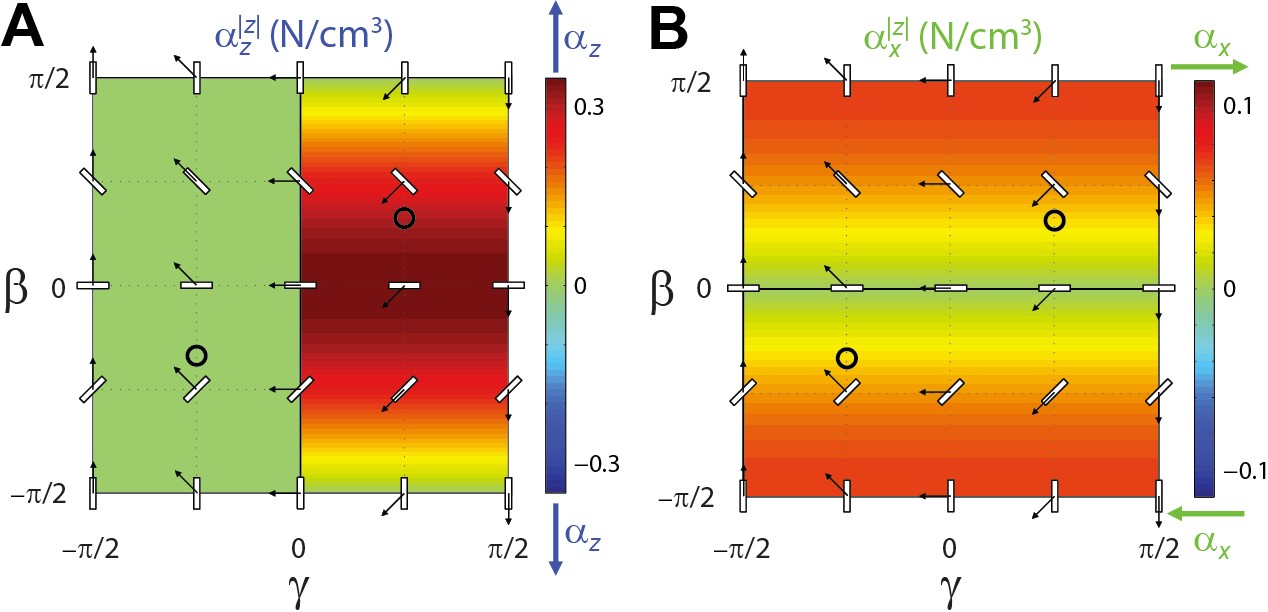}
\end{center}
\caption
{
Effective vertical ($\alpha_z^{|z|}(\beta, \gamma)$, left) and horizontal ($\alpha_{z,x}^{|z|}(\beta, \gamma)$, right) stresses per unit depth as a function of attack angle $\beta$ and intrusion angle $\gamma$ for loosely packed poppy seeds, calculated from eq.~\ref{AlphaProjected} using $\alpha_z(0,sgn(\dot{z})\pi/2)$ and $\alpha_x(\pi/2,0)$ measured in experiment (Fig.~2, C and D). See Fig.~2, A and B for schematic of the experiment and definition of variables.
}
\label{AlphaKzLPpoppy}
\end{figure}

\begin{figure}[h!]
\begin{center}
	 \includegraphics[width=3.3in]{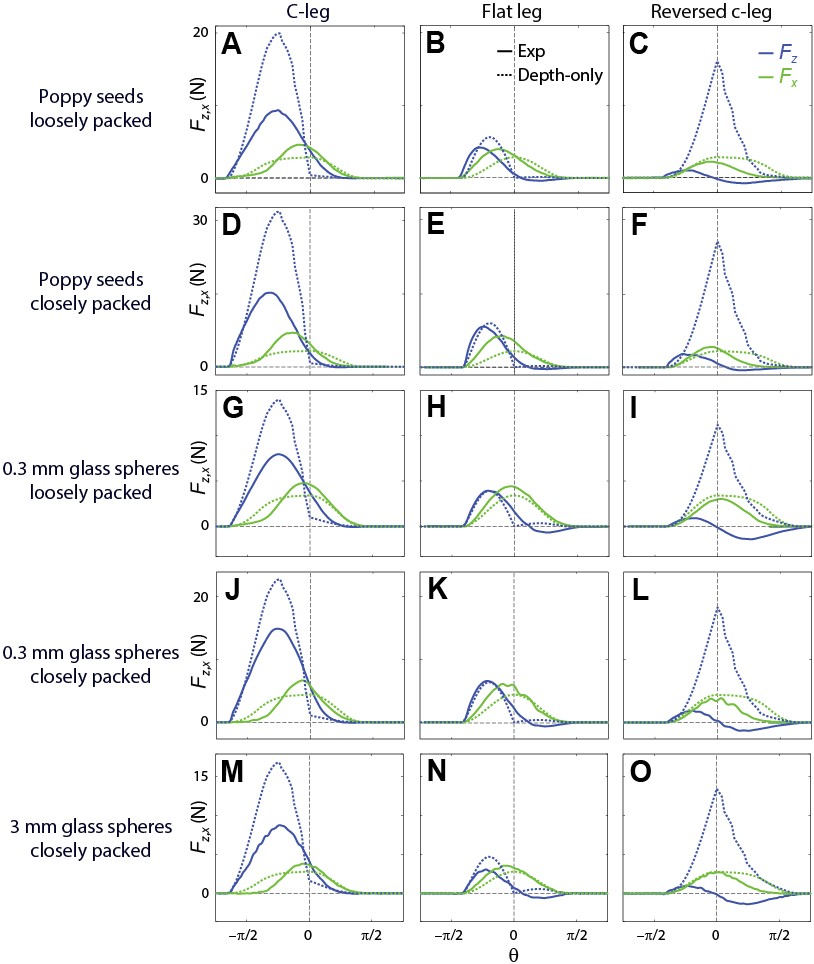}
\end{center}
\caption
{
Net lift $F_z$ (blue) and thrust $F_x$ (green) versus leg angle $\theta$ on the three model legs for all media tested. Solid curves: experimental measurements. Dotted curves: predictions from the depth-only force models. See Fig.~3 A to C for schematic of the experiment and definition of variables.
}
\label{ForceVsThetaKz}
\end{figure}

In previous studies of legged locomotion on granular media (16, 21, 37), due to the lack of the resistive force model, the forces on a complex intruder were estimated from the depth-only force models, using the flat-plate approximation: The vertical force on a leg element of area $dA_s$ and arbitrary $\beta$ and $\gamma$ was approximated by that on a horizontal leg element whose area was the element area projected into the horizontal plane (16, 21):
\begin{equation}
dA_{zs} = |\textrm{cos}\beta|dA_s
\end{equation}
Similarly, the horizontal force on the leg element was approximated by that on a vertical leg element whose area was the element area projected into the vertical plane (37):
\begin{equation}
dA_{zs} = |\textrm{sin}\beta|dA_s
\end{equation}

The effective stresses per unit depth calculated from these depth-only force models using flat plate approximation were then:
\begin{equation}
\begin{cases}
\alpha_z^{|z|}(\beta, \gamma) = \alpha_z(0, \textrm{sgn}(\dot{z})\pi/2)|\textrm{cos}\beta|\\
\alpha_x^{|z|}(\beta, \gamma) = \alpha_x(\pi/2, 0)|\textrm{sin}\beta|
\end{cases}
\label{AlphaProjected}
\end{equation}

\begin{figure}[b!]
\begin{center}
	 \includegraphics[width=3.3in]{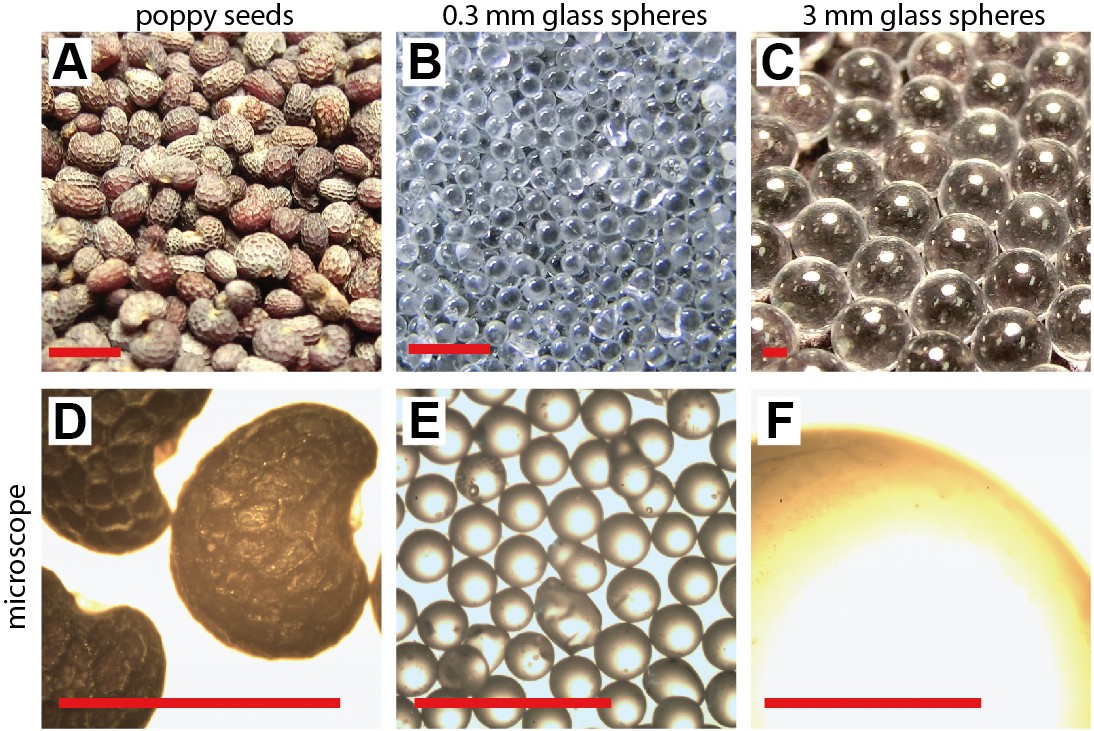}
\end{center}
\caption
{
The three granular media tested in our study. Top: regular images. Bottom: microscope images. The length of each scale bar is $1$~mm. Photo credit of (A) and (C): Sarah Sharpe.
}
\label{GranularMedia}
\end{figure}

\begin{figure}[b!]
\begin{center}
	 \includegraphics[width=3.3in]{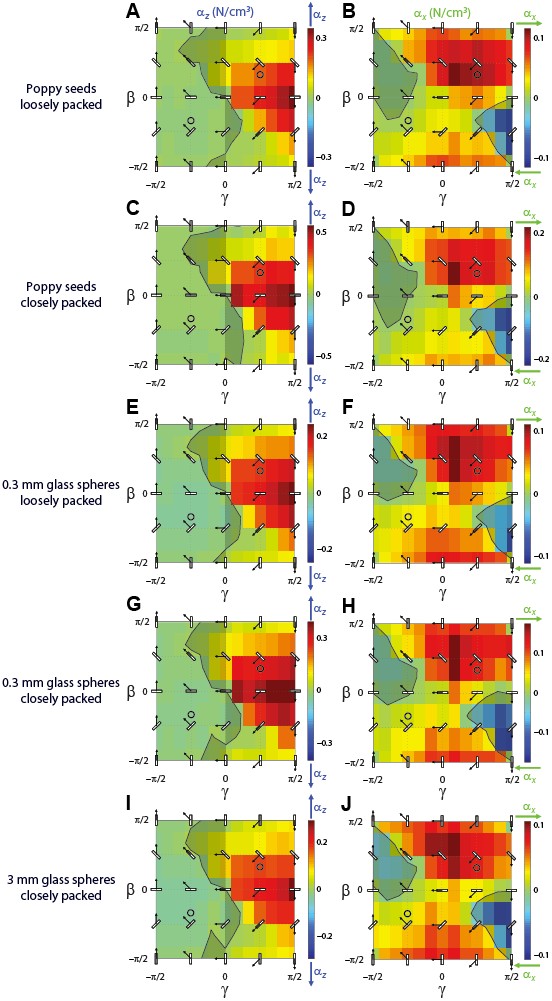}
\end{center}
\caption
{
Vertical ($\alpha_z^{|z|}(\beta, \gamma)$, left) and horizontal ($\alpha_{z,x}^{|z|}(\beta, \gamma)$, right) stresses per unit depth as a function of attack angle $\beta$ and intrusion angle $\gamma$ for all media tested. (A) and (B) are reproduced from Fig.~2, C and D. See Fig.~2, A and B for schematic of the experiment and definition of variables.
}
\label{AlphaVsBetaGamma}
\end{figure}

\begin{figure}[t!]
\begin{center}
	 \includegraphics[width=3.3in]{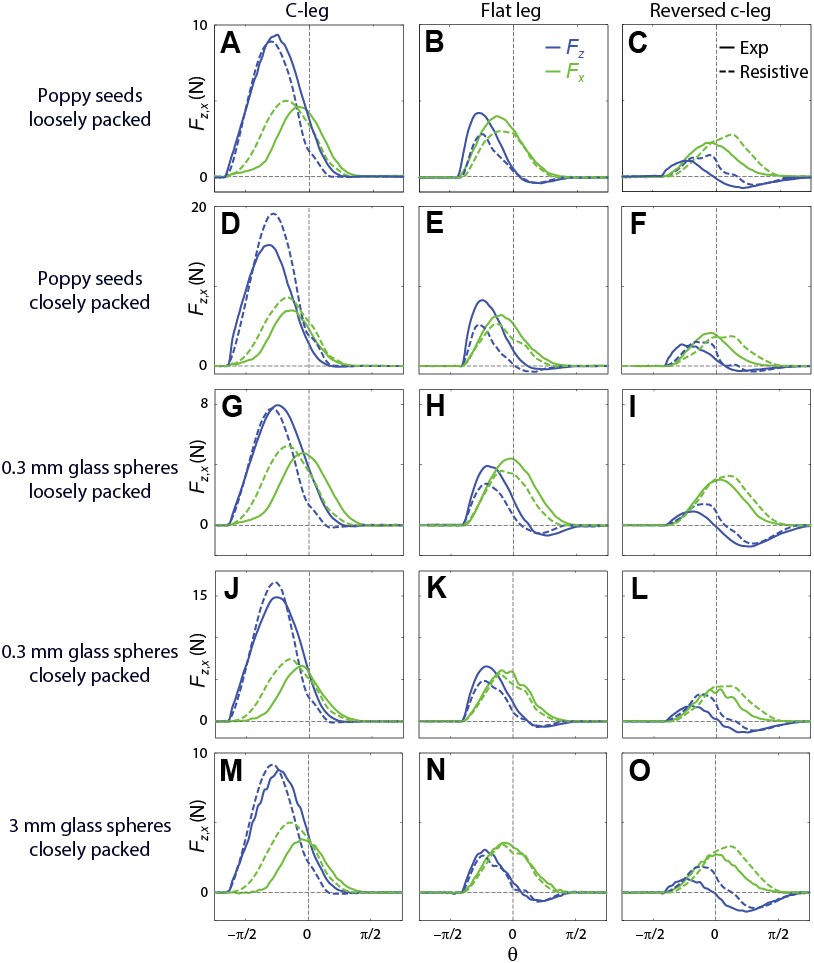}
\end{center}
\caption
{
Net lift $F_z$ (blue) and thrust $F_x$ (green) versus leg angle $\theta$ on the three model legs for all media tested. Solid curves: experimental measurements. Dashed curves: resistive force model predictions. (A) to (C) are reproduced from Fig.~3, D to F. See Fig.~3 A to C for schematic of the experiment and definition of variables.
}
\label{ForceVsTheta}
\end{figure}

Comparing the $\alpha_{z,x}^{|z|}(\beta, \gamma)$ calculated using eq.~\ref{AlphaProjected} (fig.~\ref{AlphaKzLPpoppy}) to our resistive force measurements $\alpha_{z,x}(\beta, \gamma)$ (Fig.~2, C and D) for loosely packed poppy seeds, we found that these depth-only force models did not capture most of the measured stress profiles. In particular, the depth-only vertical penetration force model overpredicted $\alpha_z(\beta, \gamma)$ for all $\beta$ and $\gamma$ except when the plate was horizontal ($\beta = 0$) and moving vertically ($\gamma = \pm \pi/2$). Contrary to experimental measurements, the $\alpha_{z,x}^{|z|}$ calculated from the depth-only models were symmetric to $\beta = 0$ (horizontal orientation), and $\alpha_z^{|z|}$ (or $\alpha_x^{|z|}$) was always opposing the plate's vertical (or horizontal) velocity (the shaded regions shrank to a line).

We used these depth-only force models to calculate the net lift $F_z$ and thrust $F_x$ versus leg angle $\theta$ for the three model legs rotated through all media tested, by the integration of stresses over the legs (Eq.~3) using $\alpha_{z,x}^{|z|}(\beta, \gamma)$. We found that they not only significantly overpredicted peak $F_{z}$ (by up to $1400\%$) but also erroneously predicted similar peak $F_{z,x}$ for the c-leg and reversed c-leg (fig.~\ref{ForceVsThetaKz}).


\begin{flushleft}
2. Generality of resistive force model for a variety of granular media
\end{flushleft}

To test the generality of our resistive force model and provide a database for future studies (we provide the force measurements in this section in Additional Data Table S5), we performed resistive force measurements for three granular media---poppy seeds, $0.3$~mm glass spheres, and $3$~m glass spheres (fig.~\ref{GranularMedia}), which have different particle size, shape, density, and friction (measured by angle of repose) (table~\ref{GranularMediaUsed}). We prepared these granular media into well-defined compactions which affected stresses (21, 38). We prepared poppy seeds and $0.3$~mm glass spheres into both a loosely packed (LP) and a closely packed (CP) states, and prepared $3$~m glass spheres into a closely packed (CP) state (table~\ref{GranularMediaUsed}).



\begin{table*}[h!]
\caption
{
Physical properties of the three granular media (in different compactions) for which resistive forces were measured. *Angle of repose for $3$~mm glass spheres courtesy of Sarah Sharpe.
}
\begin{center}
\scalebox{0.9}{
\begin{tabular}{ | c | c | c | c | c | c | }
\hline granular medium & particle diameter (mm) & particle material density (g/cm$^3$) & compaction & volume fraction & angle of repose ($^\circ$)\\
\hline \multirow{2}{*}{poppy seeds} & \multirow{2}{*}{$0.7 \pm 0.2$} & \multirow{2}{*}{1.1} & loosely packed (LP) & 0.58 & 36\\
&&& closely packed (CP) & 0.62 & 47\\
\hline \multirow{2}{*}{$0.3$~mm glass spheres} & \multirow{2}{*}{$0.27 \pm 0.04$} & \multirow{2}{*}{2.5} & loosely packed (LP) & 0.58 & 25\\
&&& closely packed (CP) & 0.62 & 35\\
\hline $3$~mm glass spheres & $3.2 \pm 0.2$ & 2.6 & closely packed (CP) & 0.63 & 21*\\
\hline
\end{tabular}}
\end{center}
\label{GranularMediaUsed}
\end{table*}





Despite differences in magnitudes and fine features, we observed similar profiles of the vertical and horizontal stresses per unit depth $\alpha_{z,x}(\beta, \gamma)$ for all media tested (fig.~\ref{AlphaVsBetaGamma}). We provide the $\alpha_{z,x}(\beta, \gamma)$ data for all media in Additional Data Table S5 (separate file in Microsoft Excel format).

For all media tested, our resistive force model predicted the net lift $F_z$ and thrust $F_x$ versus leg angle $\theta$ on the three model legs rotated through granular media (fig.~\ref{ForceVsTheta}). Compared with predictions from the depth-only force models (fig.~\ref{AlphaKzLPpoppy}), our resistive force model had a significant improvement in accuracy.


\begin{flushleft}
3. Scaling routine for easy use of the resistive force model
\end{flushleft}

To provide a means for practical use of our resistive force model and comparison with new theories of dense granular flow (32), we performed a fitting approximation to the stress per unit depth data $\alpha_{z,x}$ for all media tested, and developed a scaling routine based on the data fits.

We first performed a discrete Fourier transform of the $\alpha_{z,x}(\beta, \gamma)$ data (fig.~\ref{AlphaVsBetaGamma}) over $-\pi/2 \leq \beta \leq \pi/2$ and $-\pi \leq \gamma \leq \pi$ to obtain a fitting function. We examined the Fourier coefficients and found that the $\alpha_{z,x}(\beta, \gamma)$ data of all media tested could be well approximated by the following fits:
\begin{equation}
\begin{cases}
\begin{aligned}
\alpha_z^{\textrm{fit}}(\beta, \gamma) = \sum_{m = -1}^{1} \sum_{n = 0}^{1} [A_{m,n}cos2\pi(\frac{m\beta}{\pi} + \frac{n\gamma}{2\pi})\\ + B_{m,n}sin2\pi(\frac{m\beta}{\pi} + \frac{n\gamma}{2\pi})]\\
\alpha_x^{\textrm{fit}}(\beta, \gamma) = \sum_{m = -1}^{1} \sum_{n = 0}^{1} [C_{m,n}cos2\pi(\frac{m\beta}{\pi} + \frac{n\gamma}{2\pi})\\ + D_{m,n}sin2\pi(\frac{m\beta}{\pi} + \frac{n\gamma}{2\pi})]
\end{aligned}
\end{cases}
\label{FourierExpansion}
\end{equation}
using nine zeroth- and first-order terms (whose magnitudes are larger than $0.05 A_{0,0}$) (table~\ref{FourierCoeffs}):
\begin{equation}
M =
\begin{pmatrix}
A_{0,0} & A_{1,0} & B_{1,1} & B_{0,1} & B_{-1,1} & C_{1,1} & C_{0,1} & C_{-1,1} & D_{1,0}
\end{pmatrix}^T
\label{MatrixDefinition}
\end{equation}


\begin{figure}[b!]
\begin{center}
	 \includegraphics[width=3.3in]{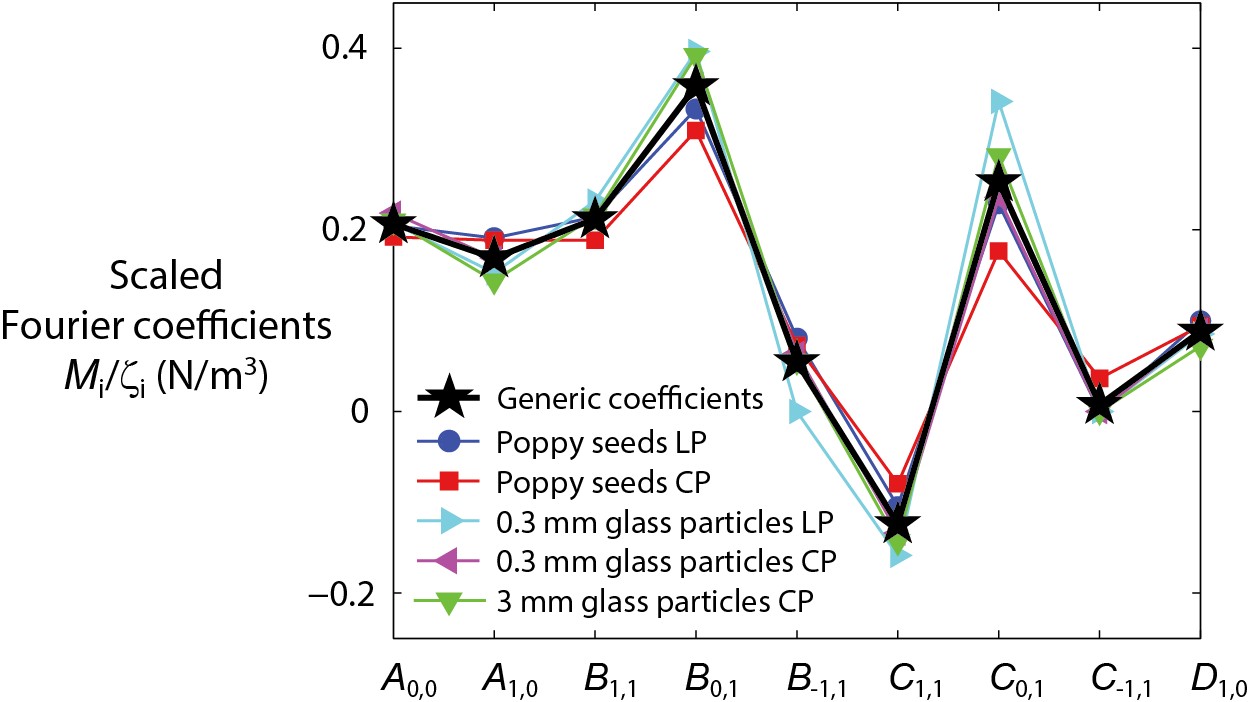}
\end{center}
\caption
{
Scaled Fourier coefficients $M_i / \zeta_i$ of all media tested (thin colored curves) can be approximated by a generic coefficient curve $M_0$ (thick black curve).}
\label{FourierCoeffsScaling}
\end{figure}

\begin{table*}[h!]
\caption
{
Zeroth- and first-order Fourier coefficients $M$ (in N/cm$^3$) for all media tested. These coefficients are $> 0.05 A_{0,0}$. $M_0$ is a generic coefficient curve to which the $M$ for each granular medium can be collapsed onto by division of a scaling factor $\zeta$  ($\zeta =$ 1 for $M_0$). We choose the magnitudes of $M_0$ such that for all media tested the values of $\zeta$ are nearly the same as the values of $\alpha_z^{\textrm{fit}}(0, \pi/2)$ (this becomes useful later in the scaling routine). CP and LP indicates closely packed and loosely packed states.
}
\begin{center}
\scalebox{0.97}{
\begin{tabular}{ | c | c | c | c | c | c | c | }
\hline granular medium & \multicolumn{2}{|c|}{poppy seeds} & \multicolumn{2}{|c|}{$0.3$~mm glass spheres} & $3$~mm glass spheres & generic coefficients\\
\hline compaction & LP & CP & LP & CP & CP & n/a\\
\hline matrix notation & $M_{\textrm{poppyLP}}$ & $M_{\textrm{poppyCP}}$ & $M_{\textrm{0.3mmLP}}$ & $M_{\textrm{0.3mmCP}}$ & $M_{\textrm{3mmCP}}$ & $M_0$\\
\hline $A_{0,0}$ & 0.051 & 0.094 & 0.040 & 0.081 & 0.045 & 0.206\\
\hline $A_{1,0}$ & 0.047 & 0.092 & 0.030 & 0.063 & 0.031 & 0.169\\
\hline $B_{1,1}$ & 0.053 & 0.092 & 0.045 & 0.078 & 0.046 & 0.212\\
\hline $B_{0,1}$ & 0.083 & 0.151 & 0.077 & 0.133 & 0.084 & 0.358\\
\hline $B_{-1,1}$ & 0.020 & 0.035 & 0 & 0.024 & 0.012 & 0.055\\
\hline $C_{1,1}$ & $-$0.026 & $-$0.039 & $-$0.031 & $-$0.050 & $-$0.031 & $-$0.124\\
\hline $C_{0,1}$ & 0.057 & 0.086 & 0.066 & 0.087 & 0.060 & 0.253\\
\hline $C_{-1,1}$ & 0 & 0.018 & 0 & 0 & 0 & 0.007\\
\hline $D_{1,0}$ & 0.025 & 0.046 & 0.017 & 0.033 & 0.015 & 0.088\\
\hline scaling factor $\zeta$ & 0.248 & 0.488 & 0.194 & 0.371 & 0.214 & 1\\
\hline
\end{tabular}}
\end{center}
\label{FourierCoeffs}
\end{table*}

By symmetry, $\alpha_z(\beta \leq 0, \pi/2) = \alpha_z(\beta \geq 0, \pi/2)$, and $\alpha_x(\beta \leq 0, \pi/2) = -\alpha_x(\beta \geq 0, \pi/2)$. However, the data slightly deviated from this equality because the initial positions of the plate were close to one of the boundaries of the container. Therefore, before the Fourier transform, we averaged the raw data for $\gamma = \pm \pi/2$ to restore symmetry by $\alpha_z(\beta \leq 0, \pi/2) = \alpha_z(\beta \geq 0, \pi/2) = \frac{1}{2}[\alpha_z(\beta \leq 0, \pi/2) + \alpha_z(\beta \geq 0, \pi/2)]$. The raw data were also not uniformly sampled in the $\gamma$ direction; we found that this only resulted in small errors in data fitting.

We found that for all media tested (denoted by $i =$ poppyLP, poppyCP, $0.3$mmLP, $0.3$mmCP, and $3$mmCP), the Fourier coefficients $M_i$ could be collapsed onto a generic coefficient curve, $M_0$, by dividing $M_i$ by a scaling factor $\zeta_i$ (table~\ref{FourierCoeffs}, fig.~\ref{FourierCoeffsScaling}):
\begin{equation}
M_{i}/\zeta_i \approx M_0
\label{FourierCoeffsScaling}
\end{equation}

\begin{figure}[t!]
\begin{center}
	 \includegraphics[width=3.3in]{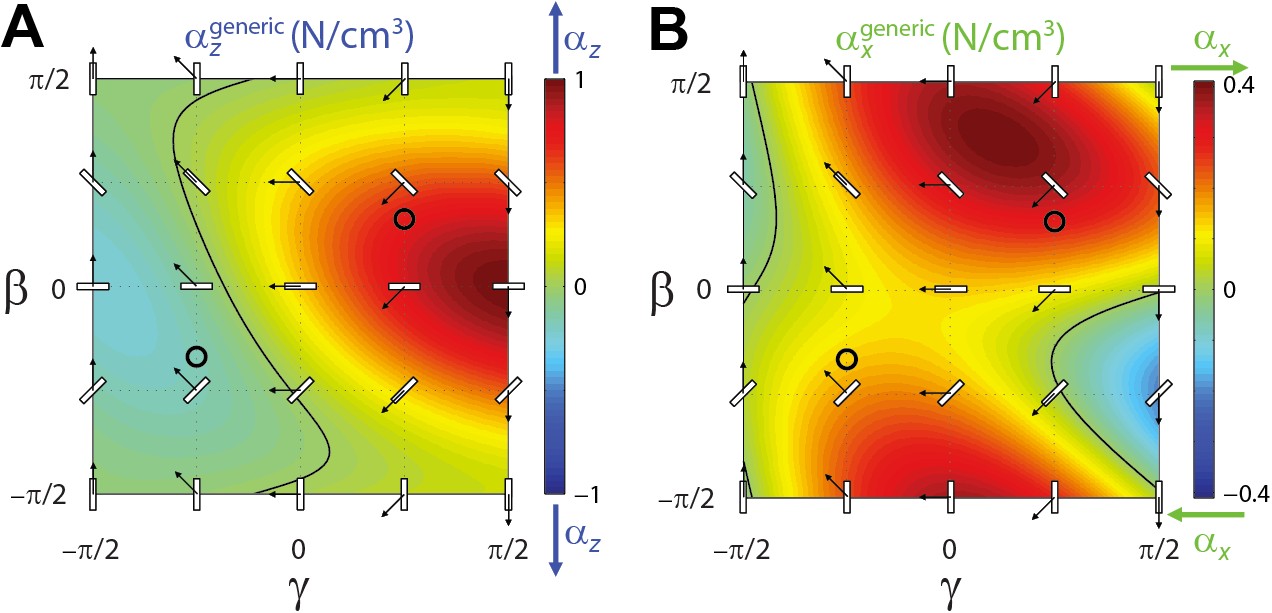}
\end{center}
\caption
{
Generic stress (per unit depth) profiles $\alpha_{z,x}^{\textrm{generic}}(\beta, \gamma)$ for all media tested. See Fig.~2, A and B for schematic of the experiment and definition of variables.
}
\label{AlphaVsBetaGammaGeneric}
\end{figure}

This enabled us to scale stresses per unit depth ($\alpha_{z,x}$) and thus forces ($f_{z,x}$ and $F_{z,x}$) for all media tested. By eq.~\ref{FourierExpansion}, using the generic coefficient curve $M_0$ ($\zeta = 1$) from table~\ref{FourierCoeffs}, we calculated generic stress (per unit depth) profiles $\alpha_{z,x}^{\textrm{generic}}(\beta, \gamma)$ (fig.~\ref{AlphaVsBetaGammaGeneric}). We found that, by multiplication by the scaling factor $\zeta$, these generic stress profiles well approximated the measured $\alpha_{z,x}(\beta, \gamma)$ (fig.~\ref{AlphaVsBetaGamma}) for all media tested:
\begin{equation}
\alpha_{z,x}(\beta, \gamma) \approx \zeta\alpha_{z,x}^{\textrm{generic}}(\beta, \gamma)
\label{StressProfileScaling}
\end{equation}


\begin{figure}[b!]
\begin{center}
	 \includegraphics[width=3.3in]{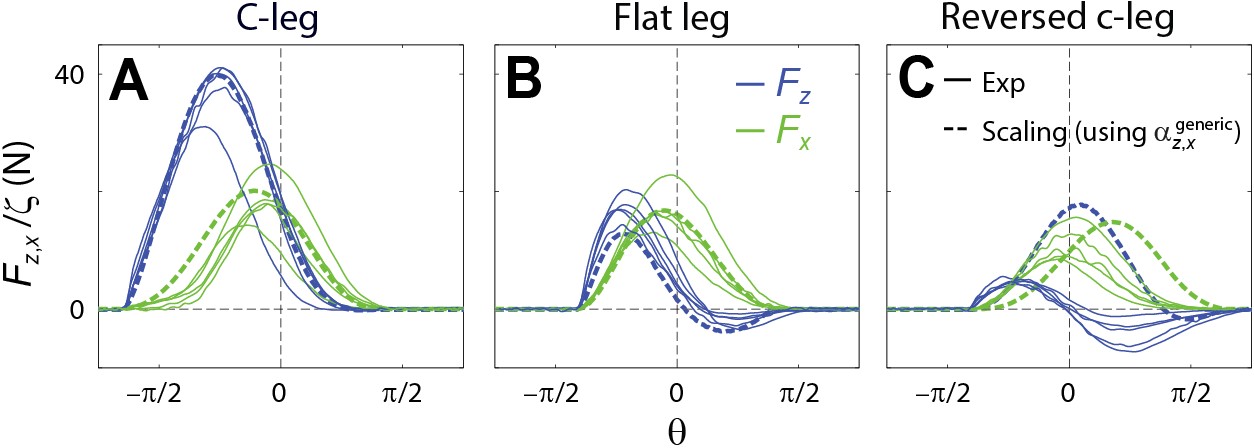}
\end{center}
\caption
{
The measured net forces scaled by the measured scaling factors $F_{z,x}(\theta)/\zeta$ for all media tested (thin curves), in comparison with the generic profiles of net forces $F_{z,x}^{\textrm{generic}}(\theta)$ (thick curves) calculated from the generic stress profiles $\alpha^{\textrm{generic}}_{z,x}$ (fig.~\ref{AlphaVsBetaGammaGeneric}). See Fig.~3 A to C for schematic of the experiment and definition of variables.
}
\label{ForceVsThetaScaled}
\end{figure}

Using Eq.~2 and $\alpha_{z,x}^{\textrm{generic}}(\beta, \gamma)$, we calculated the generic force profiles $F_{z,x}^{\textrm{generic}}(\theta)$ on the three model legs rotated through granular media (fig.~\ref{ForceVsThetaScaled}). We found that in all media tested and for both the c-leg and the flat leg, these generic force profiles captured the measured $F_{z,x}$ scaled by the scaling factors $\zeta$:
\begin{equation}
F_{z,x}^{\textrm{generic}}(\theta) \approx F_{z,x}(\theta)/\zeta
\end{equation}
However, the agreement was worse for the reversed c-leg. Our resistive force model revealed that this was because by using $F_{z,x}^{\textrm{generic}}(\theta)$, stresses were significantly overestimated for the reversed c-leg's elements that reached large depths.


We further observed that the ratio of the maximal vertical stress (which occurred on a horizontal plate moving downward) between the measurements ($\alpha_z(0, \pi/2)$) and fits ($\alpha_z^{\textrm{fit}}(0, \pi/2)$) was similar for all media tested (table~\ref{VerticalPenetrationStressPerUnitDepth}):
\begin{equation}
\chi = \alpha_z(0, \pi/2)/\alpha_z^{\textrm{fit}}(0, \pi/2) = 1.26 \pm 0.14
\end{equation}
where $\chi$ is in mean $\pm$~s.d.

\begin{table*}[h!]
\caption
{
Comparison of the measurements and fits of maximal vertical stress per unit depth (in N/cm$^3$) for all media tested. CP and LP indicates closely packed and loosely packed states.
}
\begin{center}
\scalebox{1}{
\begin{tabular}{ | c | c | c | c | c | c | c | }
\hline granular medium & \multicolumn{2}{|c|}{poppy seeds} & \multicolumn{2}{|c|}{$0.3$~mm glass spheres} & $3$~mm glass spheres\\
\hline compaction & LP & CP & LP & CP & CP\\
\hline $\alpha_z(0, \pi/2)$ & 0.35 & 0.56 & 0.24  & 0.40 & 0.29\\
\hline $\alpha_z^{\textrm{fit}}(0, \pi/2)$ & 0.26 & 0.47 & 0.19  & 0.38 & 0.22\\
\hline $\chi = \alpha_z(0, \pi/2)/\alpha_z^{\textrm{fit}}(0, \pi/2)$ & 1.37 & 1.19 & 1.27  & 1.05 & 1.33\\
\hline
\end{tabular}}
\end{center}
\label{VerticalPenetrationStressPerUnitDepth}
\end{table*}

Therefore, we propose that for a sufficiently level and uniform dry granular medium composed of near-monodispersed, near-spherical, rounded particles of $\sim 0.1$ to $\sim 1$~mm in diameter, one can simply measure its maximal vertical stress $\alpha_z(0, \pi/2)$ by pushing a horizontal plate downward to infer the maximal value of the fit vertical stress $\alpha_z^{\textrm{fit}}(0, \pi/2)$:
\begin{equation}
\alpha_z^{\textrm{fit}}(0, \pi/2) = \alpha_z(0, \pi/2)/\chi \approx 0.8\alpha_z(0, \pi/2)
\end{equation}

\begin{figure}[b!]
\begin{center}
	 \includegraphics[width=3.3in]{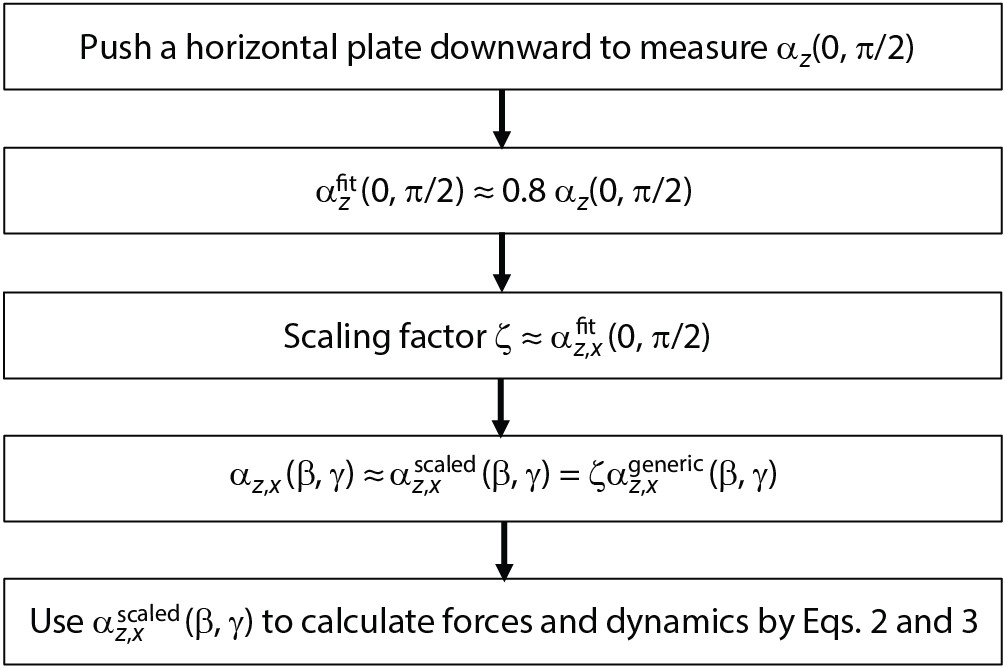}
\end{center}
\caption
{
A scaling routine to easily apply our resistive force model.
}
\label{StepsToUseOurModel}
\end{figure}

The value of $\alpha_z^{\textrm{fit}}(0, \pi/2)$ in N/m$^3$ then gives the scaling factor $\zeta$ for this granular medium, because we choose the magnitudes of the generic curve $M_0$ so that the values of $\zeta$ and $\alpha_z^{\textrm{fit}}(0, \pi/2)$ are nearly the same for all media tested:
\begin{equation}
\zeta \approx \alpha_{z,x}^{\textrm{fit}}(\beta, \gamma)
\end{equation}
Note that this equation only equates the numeric values on both sides, because $\zeta$ is dimensionless.

Then, from eq.~\ref{StressProfileScaling}, by scaling the generic stress profiles $\alpha^{\textrm{generic}}_{z,x}$ (fig.~\ref{AlphaVsBetaGammaGeneric}) by the determined scaling factor $\zeta$, we can obtain an approximation of stress profiles for this granular medium:
\begin{equation}
\alpha_{z,x}(\beta, \gamma) \approx \alpha_{z,x}^{\textrm{scaled}}(\beta, \gamma) = \zeta\alpha_{z,x}^{\textrm{generic}}(\beta, \gamma)
\label{ScaledStressProfiles}
\end{equation}

This scaling routine provides an alternative to measuring $\alpha_{z,x}$ for all $\beta$ and $\gamma$ (at the cost of model accuracy). As demonstrated by the model leg rotation experiments (fig.~\ref{ForceVsThetaScaled}) and robot locomotion experiments (see fig.~\ref{SpeedVsFreqCurv} in the next section), our scaling routine only suffers a small loss in accuracy for much of the leg morphology and stride frequencies tested. This technique can be particulary useful in a field setting, because only a single force measurement is needed.

We summarize these practical steps to use our resistive force model in fig.~\ref{StepsToUseOurModel}.



\begin{flushleft}
4. Applicability of resistive force model to natural sands
\end{flushleft}

\begin{figure}[b!]
\begin{center}
	 \includegraphics[width=3.3in]{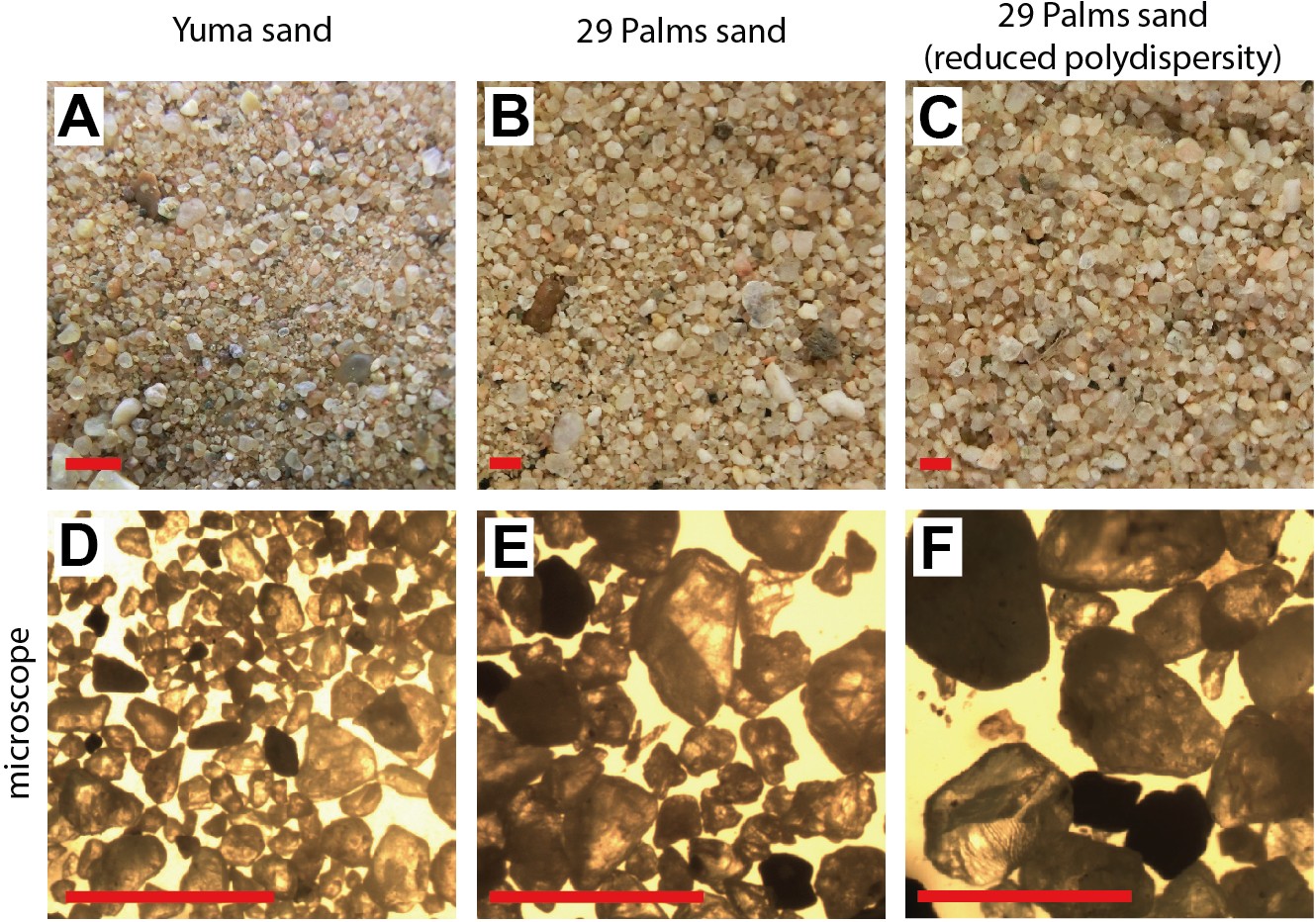}
\end{center}
\caption
{
The natural sands used to test the predictive power of our scaling routine. Top: regular images. Bottom: microscope images. The length of each scale bar is $1$~mm.
}
\label{NaturalSandFig}
\end{figure}

To test the predictive power of our resistive force model for natural sands, we chose two natural sands of higher polidispersity and angularity than those of the granular media tested, and examined whether the scaling routine (fig.~\ref{StepsToUseOurModel}) could predict the net forces $F_{z,x}(\theta)$ on the three model legs during rotation through the natural sands.

The two natural sands, referred to as ``Yuma sand'' and ``29 Palms sand'' (table~\ref{NaturalSandTable}, fig.~\ref{NaturalSandFig}), were collected from the Mojave Desert in the southwest United States, one from Yuma, Arizona and the other from 29 Palms, California. The Yuma sand had most particles (70\% by mass) in the 0.06--0.3~mm particle size range, and the 29 Palms sand had most particles (91\% by mass) in the 0.3--3~mm particle size range. Both naturals sands were fluidized by the fluidized bed before each force measurement was performed. We further tested modified 29 Palms sand with reduced polydispersity to examine the effect of polydispersity on model accuracy.



\begin{table*}[h!]
\caption
{
Particle size distribution of the natural sands tested.
}
\begin{center}
\scalebox{1}{
\begin{tabular}{ | c | c | c | c | c | c | }
\hline natural sand & particle diameter (mm) & mass percentage (\%)\\
\hline \multirow{3}{*}{Yuma sand}
& $<$ 0.06 & 2\\
& 0.06--0.3 & 68\\
& 0.3--3.0 & 17\\
& $>$3.0 & 13\\
\hline \multirow{5}{*}{29 Palms sand}
& $<$ 0.3 & 5\\
& 0.3--0.6 & 36\\
& 0.6--0.7 & 55\\
& 0.7--3.0 & 3\\
& $>$3.0 & 1\\
\hline 29 Palms sand (reduced polydispercity) & 0.6--0.7 & 100\\
\hline
\end{tabular}}
\end{center}
\label{NaturalSandTable}
\end{table*}

\begin{figure}[t!]
\begin{center}
	 \includegraphics[width=3.3in]{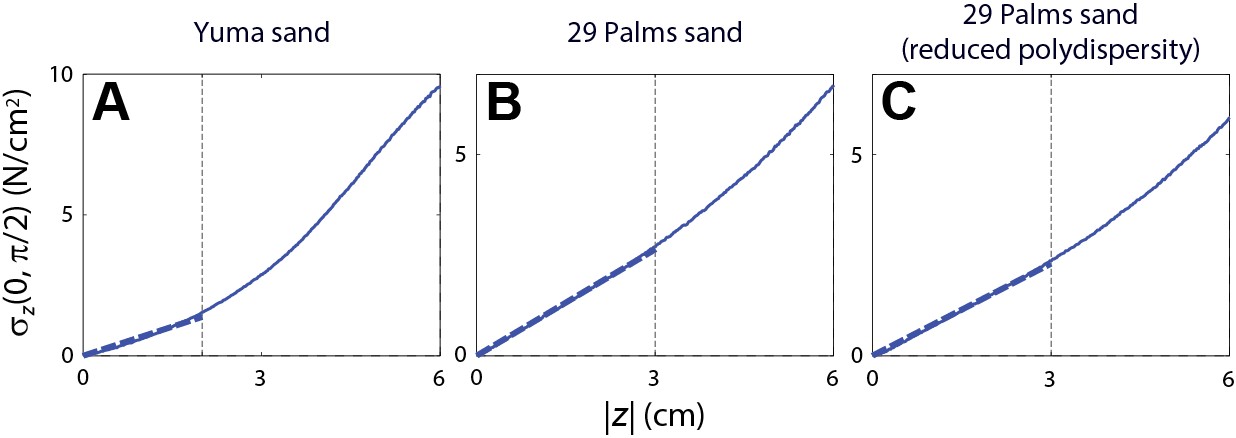}
\end{center}
\caption
{
Maximal vertical stress $\sigma_z(0, \pi/2)$ versus depth $|z|$ measured in the natural sands using a plate element horizontally oriented and penetrating vertically downward ($\beta = 0, \gamma = \pi/2$). Dashed lines are linear fits with zero intercept to the data in the linear regime at shallow depths. See Fig.~2, A and B for schematic of the experiment and definition of variables.
}
\label{NaturalSandPenetration}
\end{figure}

\begin{figure}[b!]
\begin{center}
	 \includegraphics[width=3.3in]{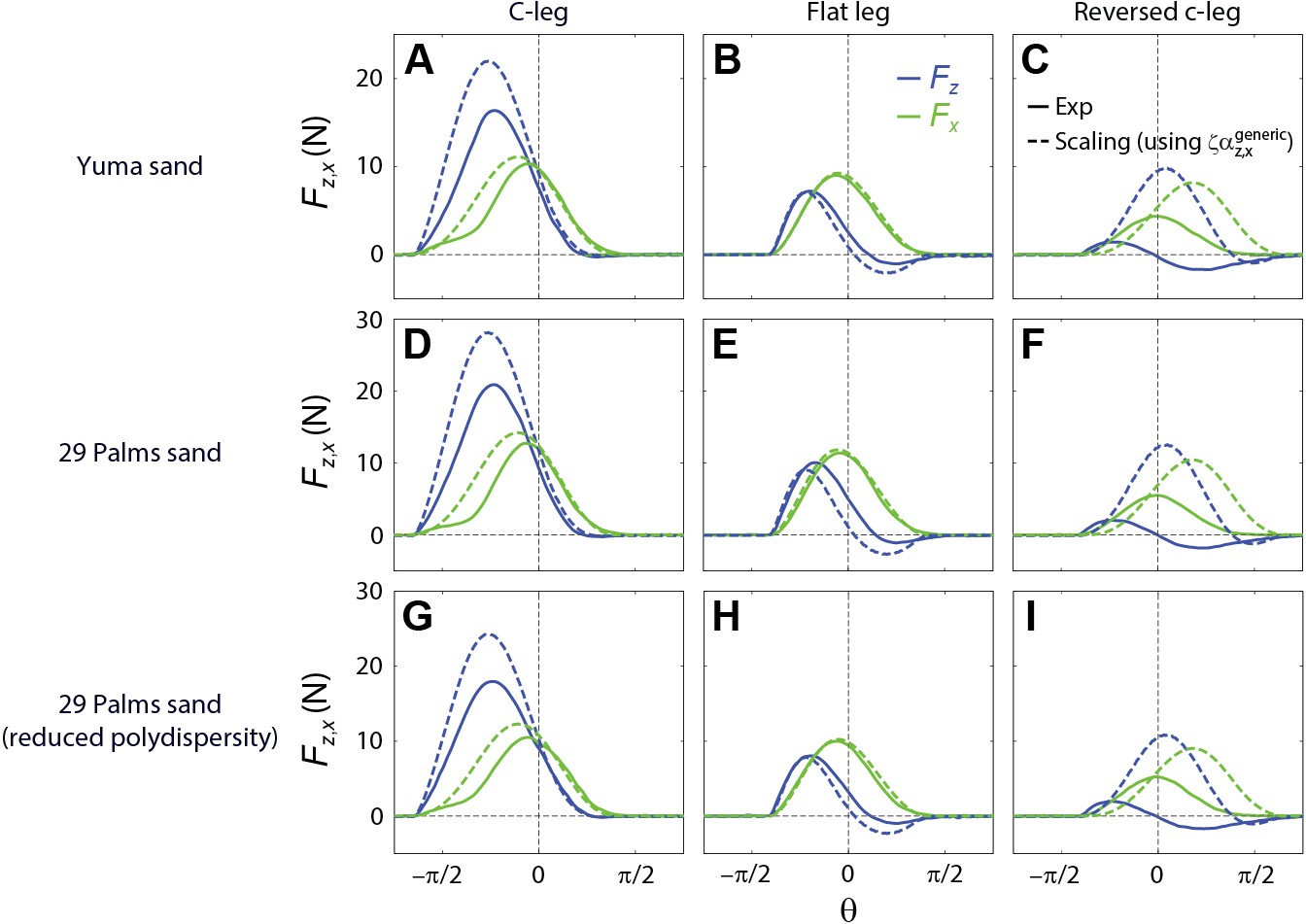}
\end{center}
\caption
{
Net lift $F_z$ (blue) and thrust $F_x$ (green) versus leg angle $\theta$ on the three model legs for the natural sands tested. Solid curves: experimental measurements. Dashed curves: resistive force model predictions using the scaling routine (fig.~\ref{StepsToUseOurModel}). See Fig.~3 A to C for schematic of the experiment and definition of variables.
}
\label{NaturalSandRotation}
\end{figure}

During downward penetration ($\gamma = \pi/2$) into both the Yuma and 29 Palms sands, the vertical stress $\sigma_z$ on a horizontal plate ($\beta = 0$) increased nearly proportionally to depth $|z|$ at low enough depths (fig.~\ref{NaturalSandPenetration}, solid curves), similar to observations in the granular media tested. However, for both the natural sands, $\sigma_z$ vesus $|z|$ displayed nonlinearity as depth further increased sooner than observed for the granular media tested. This was because both natural sands had larger internal stresses (larger $\alpha_{z,x}$ magnitudes) and likely suffered boundary effects at much shallower depths as compared to the granular media tested. In addition, for the Yuma sand which had more particles in the 0.06--0.3~mm particle size range, this nonlinearity due to boundary effects was more pronounced. Therefore, to minimize possible errors from boundary effects, we perform linear fits with zero intercept to the $\sigma_z$ versus $|z|$ data (fig.~\ref{NaturalSandPenetration}, dashed curves) over the linear regime (0--2~cm for the Yuma sand; 0--3~cm for the 29 Palms sand) to obtain $\alpha_z(0, \pi/2)$ (slopes of dashed lines) for both the natural sands. This gave us their scaling factors $\zeta$ and estimates of their stress profiles from $\alpha_{z,x}^{\textrm{scaled}}(\beta, \gamma) = \zeta\alpha_{z,x}^{\textrm{generic}}(\beta, \gamma)$ (eq.~\ref{ScaledStressProfiles}).


For the three model legs rotated through both the Yuma and 29 Palms sands, both the net lift $F_z$ and thrust $F_x$ versus leg angle $\theta$ displayed asymmetric profiles similar to those observed for the granular media tested (fig.~\ref{NaturalSandRotation}, solid curves). Using the $\alpha_z(0, \pi/2)$ obtained from the vertical penetration experiment (fig.~\ref{NaturalSandPenetration}), our scaling routine significantly overpredicted $F_{z,x}(\theta)$ on the c-leg for both the Yuma and 29 Palms sands (both by 35\%). Nevertheless, the shape of model predictions were similar to experimental observations (fig.~\ref{NaturalSandRotation}, dashed curves). The model accuracy was best for the flat leg, and worst for the reversed c-leg.


\begin{figure}[b!]
\begin{center}
	 \includegraphics[width=3.3in]{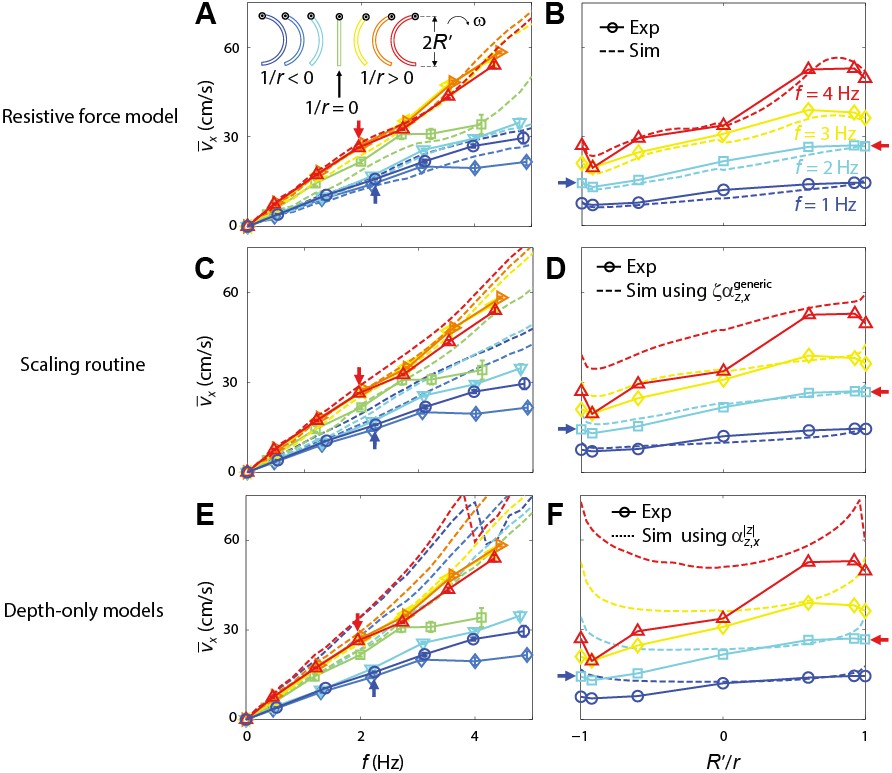}
\end{center}
\caption
{
Average forward speed $\overline{v}_x$ of the robot as a function of stride frequency $f$ (left) and leg curvature $1/r$ (right) on loosely packed poppy seeds. Solid curves: experimental measurements. Dashed curves: simulation predictions. Top: using the measured stresses ($\alpha_{z,x}$). Middle: using the scaled generic stress profiles ($\alpha^{\textrm{scaled}}_{z,x} = \zeta\alpha^{\textrm{generic}}_{z,x}$). Bottom: using stresses from the depth-only force models ($\alpha^{|z|}_{z,x}$). The drop of $\overline{v}_x$ vs. $f$ at near $R'/r = 1$ using $\alpha^{|z|}_{z,x}$ was due to pitch instability of the robot in simulation (the robot became upside down). (A) and (B) are reproduced from Fig.~4, C and D. See Fig.~4 A and B for representative runs of the experiment and definition of variables.
}
\label{SpeedVsFreqCurv}
\end{figure}

There are two major differences between the natural sands and the granular media tested: The natural sands have higher polydispersity, and are also less spherical and more angular. To probe where the overprediction (of forces on the c-leg) stemmed from, we further tested the 29 Palms sand with reduced polydispersity (obtained by sieving the sand to obtain only the 0.6--0.7~mm particles). We found that while both $\alpha_z(0, \pi/2)$ and $F_{z,x}(\theta)$ dropped by 14\% (figs.~\ref{NaturalSandPenetration} and \ref{NaturalSandRotation}), the model overprediction of forces on the c-leg remained the same (35\%). This suggested that the non-spherical and angular shape of the natural sand particles, rather than their higher polydispersity, was likely the cause of the model overprediction.



\begin{flushleft}
5. Comparison of model accuracy in predicting legged locomotion
\end{flushleft}

To evaluate the accuracy of our scaling routine in predicting legged locomotion, we used our multibody dynamic simulation to simulate the robot's movement on loosely packed poppy seeds using the scaled generic stress profiles ($\alpha^{\textrm{scaled}}_{z,x} = \zeta\alpha^{\textrm{generic}}_{z,x}$). We calculated the robot's average forward speed $\overline{v}_x$ versus stride frequency $f$ and leg curvature $1/r$ (fig.~\ref{SpeedVsFreqCurv}), and compare it with experimental measurements and resistive force model predictions (without using the scaling routine). As a comparison, we also calculated $\overline{v}_x(f, 1/r)$ using $\alpha^{|z|}_{z,x}$ obtained from the depth-only force models (fig.~\ref{AlphaKzLPpoppy}).


\begin{figure}[b!]
\begin{center}
	 \includegraphics[width=1.8in]{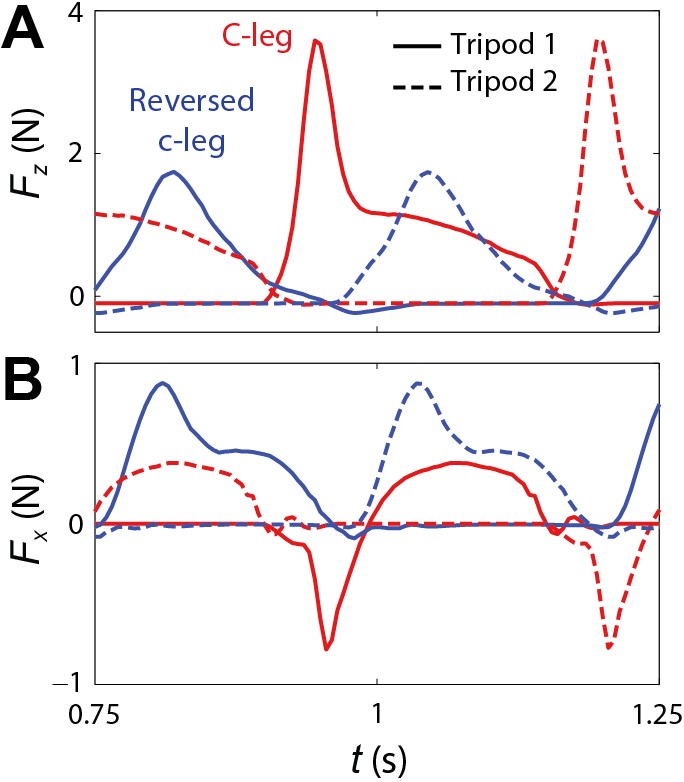}
\end{center}
\caption
{
Vertical ($F_z$, top) and horizontal ($F_x$, bottom) ground reaction forces versus time $t$ on both tripods of the robot during locomotion on loosely packed poppy seeds, calculated from the simulation using the resistive force model. Solid and dashed curves correspond to forces on the two tripods of legs, respectively. Results are for the two representative runs shown in Fig.~4B. Note that these ground reaction forces do not include those on the robot body (which were also calculated but not plotted here).
}
\label{GroundReactionForce}
\end{figure}

We found that the simulation using scaled generic stress profiles ($\zeta\alpha^{\textrm{generic}}_{z,x}$) only suffered a small loss in accuracy as compared to that using the measured $\alpha_{z,x}$. By contrast, the simulation using stresses $\alpha^{|z|}_{z,x}$ from the depth-only force models suffered a larger loss in accuracy (by up to $150\%$ error), and in particular, erroneously predicted that the robot moved at similar speeds using both c-legs and reversed c-legs.

In addition to the predictions locomotor kinematics, our resistive force model and multibody dynamic simulation also allowed us to calculate ground reaction forces (fig.~\ref{GroundReactionForce}) during locomotion on granular media (which would be otherwise difficult to measure), in a similar fashion as calculations of lift, drag, thrust, traction, and resistance in studies of vehicle mobility in fluids and terramechanics studies. Notice the larger vertical ground reaction forces produced by the C-legs relative to those by the reversed c-legs. This allowed the robot to maintain higher lift above the granular medium, reducing drag on the belly.



\clearpage

\begin{flushleft}
\textbf{Movie S1}

\textbf{\url{http://crablab.gatech.edu/pages/movies/files/Terradyn_s1.mov}}
\end{flushleft}

Plate element intrusion experiments. Left: Intrusion for representative attack angle and intrusion angle $(\beta, \gamma) = (\pi/6 ,\pi/4)$. Right: Extraction for representative $(\beta, \gamma) = -(\pi/6 ,\pi/4)$. Top: Videos of the plate moving in a granular medium (loosely packed poppy seeds). Bottom:  Vertical ($\sigma_z$, blue curve) and horizontal ($\sigma_x$, green curve) stresses versus depth $|z|$. Videos were taken with the plate at the container boundary for illustration purpose only. Forces were measured with the plate in the middle of the container, far away from the boundary. The red lines originating from the center of the top panels indicate granular forces on the plate element. The plate was moved at $1$~cm/s. See Fig.~2, A and B for schematic of the experiment and definition of variables.

\begin{flushleft}
\textbf{Movie S2}

\textbf{\url{http://crablab.gatech.edu/pages/movies/files/Terradyn_s2.mov}}
\end{flushleft}

Leg rotation experiments. Left: C-leg. Right: Reversed c-leg. Top: Videos of the leg rotating through a granular medium (loosely packed poppy seeds). Bottom: Net lift ($\sigma_z$, blue curve) and thrust ($\sigma_x$, green curve) versus leg angle $\theta$. Videos were taken with the plate at the container boundary for illustration purpose only. Forces were measured with the plate in the middle of the container, far away from the boundary. The red lines originating from the center of the top panels indicate granular forces on the model legs. The model legs were rotated at $0.2$~rad/s (leg tip speed $\sim 1$~cm/s). Note that in the c-leg video the axle deflected slightly during rotation---this was an artifact due to the extra-long supporting rod in order for the plate to reach the boundary for illustration purpose only. For force measurements performed in the middle of the container, the supporting rod was shorter and the axle did not deflect. See Fig.~3 A to C for schematic of the experiment and definition of variables.

\begin{flushleft}
\textbf{Movie S3}

\textbf{\url{http://crablab.gatech.edu/pages/movies/files/Terradyn_s3.mov}}
\end{flushleft}

Leg rotation model calculations. Left: C-leg. Right: Reversed c-leg. Top: Videos of the leg rotating through a granular medium (loosely packed poppy seeds). Bottom: Net lift ($\sigma_z$, blue curve) and thrust ($\sigma_x$, green curve) versus leg angle $\theta$. The thick red lines originating from the center of the top panels indicate net forces on the model legs. The thin red lines on the model legs indicate element forces (on a larger scale). See Fig.~3 A to C for schematic of the experiment and definition of variables.

\begin{flushleft}
\textbf{Movie S4}

\textbf{\url{http://crablab.gatech.edu/pages/movies/files/Terradyn_s4.mov}}
\end{flushleft}

Robot (body length $= 13$ cm, body mass $m = 150$ g) running on a granular medium (loosely packed poppy seeds) using c-legs and stride frequency $f = 5$ Hz. Video is played in real time.

\begin{flushleft}
\textbf{Movie S5}

\textbf{\url{http://crablab.gatech.edu/pages/movies/files/Terradyn_s5.mov}}
\end{flushleft}

Robot experiments. Top: Using c-legs ($f = 2.0$~Hz). Middle: Using reversed c-legs ($f = 2.2$~Hz). Bottom: Forward speed $v_x$ versus time $t$. Videos are played 5 times slower than real time. Note that in the trial using reversed c-legs, the robot body orientation was reversed; we verified that this did not affect robot kinematics and speed (because the center of mass of the robot was tuned to overlap with the geometric center of the body). See Fig.~4 A for definition of variables.

\begin{flushleft}
\textbf{Movie S6}

\textbf{\url{http://crablab.gatech.edu/pages/movies/files/Terradyn_s6.mov}}
\end{flushleft}

Robot simulation. Top: Using c-legs ($f = 2.0$~Hz). Middle: Using reversed c-legs ($f = 2.2$~Hz). Bottom: Forward speed $v_x$ versus time $t$. Videos are played 5 times slower than real time. The red and green arrows on the two tripod of legs indicate the ground reaction forces on the leg elements calculated from simulation. Note that the forces on the body are not shown. See Fig.~4 A for definition of variables.


\begin{flushleft}
\textbf{Additional Data Table S5 (separate file in Microsoft Excel format)}

\textbf{\url{http://crablab.gatech.edu/pages/movies/files/Terradyn_table_s5.xls}}
\end{flushleft}
Original data of vertical ($\alpha_z$) and horizontal ($\alpha_x$) stresses per unit depth versus attack angle $\beta$ and intrusion angle $\gamma$, for all media tested. These data were used to produce Fig.~2, C and D and fig.~\ref{AlphaVsBetaGamma}.

%

\end{document}